\documentclass[11pt,a4paper]{article}

\usepackage[a4paper, margin=3.5cm]{geometry}
\usepackage[utf8]{inputenc}
\usepackage{amsmath, amsthm}
\usepackage{amsfonts}
\usepackage{amssymb}
\usepackage{graphicx}
\usepackage{color}
\usepackage{xcolor}
\usepackage[hyphens]{url}
\usepackage{hyperref}
\usepackage[round]{natbib}
\usepackage{bm}
\usepackage{paralist}
\usepackage{booktabs}
\usepackage{authblk}

\usepackage{bibunits}

\defaultbibliography{biblio}
\defaultbibliographystyle{jss2}

\newcommand{\abs}[1]{\left\vert{#1}\right\vert}

\newcommand{\ind}{\operatorname{\bf{1}}}
\newcommand{\diff}{{\,\mathrm{d}}}
\newcommand{\diag}{\operatorname{diag}}

\newcommand{\Exp}{{\mathbb{E}}}

\newcommand{\argmax}{\mathrm{argmax}}
\newcommand{\Cov}{\operatorname{Cov}}

\newcommand{\ct}{c^{(t)}}  

\newcommand{\gmst}{\mathrm{GMST'}}
\newcommand{\GEV}{\mathrm{GEV}}
\newcommand{\FDR}{\mathrm{FDR}}
\newcommand{\FWER}{\mathrm{FWER}}
\newcommand{\RL}{\mathrm{RL}}
\newcommand{\RP}{\mathrm{RP}}

\newcommand{\R}{\mathbb{R}}
\newcommand{\dev}{\mathrm{dev}}
\newcommand{\HO}{H_0^{\mathrm{ED}}}

\newcommand{\uproman}[1]{\uppercase\expandafter{\romannumeral#1}}

\theoremstyle{definition}
\newtheorem{remark}{Remark}
\newtheorem{algorithm}{Algorithm}

\begin{document}

\begin{bibunit}

\renewcommand{\thefootnote}{\fnsymbol{footnote}}
 
\title{Regional Pooling in Extreme Event Attribution Studies: an Approach Based on Multiple Statistical Testing}


\author[1]{Leandra Zanger} 
\author[1]{Axel B\"ucher \thanks{Corresponding author: axel.buecher@hhu.de}}
\author[2]{Frank Kreienkamp}
\author[2]{Philip Lorenz}
\author[2]{Jordis Tradowsky}

\affil[1]{Heinrich-Heine Universität Düsseldorf, Mathematisches Institut,\hspace{2cm} Düsseldorf, Germany}
\affil[2]{Deutscher Wetterdienst, Regionales Klimabüro Potsdam, Potsdam, Germany}

\date{\today}

\maketitle

\begin{abstract}
Statistical methods are proposed to select homogeneous locations when analyzing spatial block maxima data, such as in extreme event attribution studies.
The methods are based on classical hypothesis testing using Wald-type test statistics, with critical values obtained from suitable parametric bootstrap procedures and corrected for multiplicity. 
A large-scale Monte Carlo simulation study finds that the methods are able to accurately identify homogeneous locations, and that pooling the selected locations improves the accuracy of subsequent statistical analyses. 
The approach is illustrated with a case study on precipitation extremes in Western Europe. 
The methods are implemented in an R package that allows easy application in future extreme event attribution studies.

\medskip

\end{abstract}


\section{Introduction}

Extreme event attribution studies on precipitation extremes are typically motivated by the occurrence of an extreme event which causes major impacts such as damages to infrastructure and agriculture, or even fatalities, see, for instance, \cite{Van16, Van17, Ott18, Kre21}. A key task for attributing the event to anthropogenic climate change consists of a statistical analysis of available observational data products at the location or region of interest \citep{Phi20}.  Typically,  the observed  time period is short, often less than 100 years, which ultimately leads to large statistical uncertainties. One possibility to reduce those uncertainties is to incorporate observations from nearby locations/regions, given that their meteorological characteristics are sufficiently similar and governed by the same underlying processes to those from the region affected by an extreme event. The selection of surrounding areas for which these criteria are met can be based on expert knowledge of the meteorological characteristics and dynamics, for instance provided by experts from the national meteorological and hydrological service of the affected country, like the Deutsche Wetterdienst in Germany. The expert knowledge-based suggestion may next be assessed statistically,  which, to the best of our knowledge, has been done based on ad hoc methods in the past. In this paper, we propose profound statistical methods that can complement the expert's knowledge and which is based on statistically evaluating observational data from the past. 
Once regions with sufficiently similar characteristics of the analysed variable, e.g., the yearly maximum of daily rainfall, have been identified, the time series of all identified regions can be combined, thereby extending the available time series for the benefit of a more efficient statistical analysis.   

The building blocks for the new approach are classical Wald-type tests statistics \citep{Leh21} for testing the null hypothesis that the temporal dynamics at multiple locations of interest are the same. Unlike in the classical text-book case, and motivated by the fact that standard likelihood-based inference for extreme value distributions requires unreasonably large sample sizes for sufficient finite-sample accuracy, we employ a  parametric bootstrap device to approximate the distribution of the test statistics under the null hypothesis. This approach is motivated by results in \citealp{Lil22} for respective stationary extreme value models. Based on suitable decompositions of a global null hypothesis, we then propose to test for carefully selected sub-hypotheses, possibly after correcting the individual tests' level for multiple comparisons.

The new methods are illustrated by a large-scale Monte Carlo simulation study and by an application to the severe flooding event in Western Europe during July 2021 for which spatial pooling was applied in an attribution study following the event \citep{Kre21,Tra22}. For the benefit of researchers who would like to use this spatial pooling approach, an implementation of the method in the statistical programming environment \texttt{R} \citep{R} is publicly available as an \texttt{R} package called \texttt{findpoolreg} on GitHub \citep{findpoolreg}.

Attribution analysis of precipitation extremes is especially challenging due to short observational time series as well as their often limited spatial extend, which further complicates the detection of a trend and estimation of return periods based on the limited time series \citep[see][for a discussion on this]{Tra22}. Therefore, we will in the following present the suggested approach for a heavy rainfall event, however, the method could equally be applied to other parameters.  

The remaining parts of this paper are organized as follows. 
Section \ref{sec:mathconcept} explains the mathematical concept of the proposed methods, starting with a detailed description of the underlying model assumptions and a strategy for the detection of a possible pooling region in Section \ref{subsec:model}. It is recommended to all readers. 
In Sections \ref{subsec:ml} and \ref{subsec:waldtest}, mathematical details on the applied estimators and test statistics are given, and they may be skipped by non-expert readers. 
Next, the ideas of the bootstrap procedures that allow to draw samples from the distribution under the null hypothesis are explained. Again, this may be skipped by non-statisticians. Section \ref{subsec:combineTeststats} goes into detail about the detection strategy of possible pooling regions and the treatment of the related multiple testing problem. This section is of interest to all readers who want to apply the methods, since it provides details on how to process a set of p-values obtained from testing multiple hypotheses. 
Next, Section \ref{sec:sim} gives the results of the simulation study that was performed in order to evaluate the performance of the proposed methods. These results are of interest to all readers, and they serve as a basis for the case study conducted in Section \ref{sec:case}.
Section \ref{sec:extensions} then discusses several extensions of the proposed methods. In 
\ref{sec:regrl}, we provide a method for estimating \textit{region-wise} return periods once a pooling region has been chosen. Here, a region-wise return period of a given event is defined as the number of years that one has to wait on average until an event of the same or even greater magnitude is observed anywhere in the pooling region. 
Extensions to different model assumptions that suit e.g.\ other variables such as temperature are discussed in Section \ref{sec:extensions}. Last but not least, we come to a conclusion in Section \ref{sec:conclusion}.
Some mathematical details and further illustrations on the simulation study and the case study are postponed to a supplement.

\section{Assessing spatial homogeneities for precipitation extremes}\label{sec:mathconcept}

\subsection{A homogeneous model for precipitation extremes}\label{subsec:model}

The observational data of interest consists of annual or seasonal maximal precipitation amounts (over some fixed time duration, e.g., a day) collected over various years and at various locations (in practice, each \textit{location} may correspond to a spatial \textit{region}; we separate these two terms from the outset to avoid misunderstandings: subsequently, a \textit{region} shall be a set of locations). More precisely, we denote by $m^{(t)}_d$ the observed maximal precipitation amount in season $t$ and at location $d$, with $t=1, \dots, n$ and $d=1, \dots, D$. The location of primary interest shall be the one with index $d=1$.
Note that the choice of $d = 1$ is made for illustrative purposes only and can be replaced by any index $d \in \{1, \ldots, D\}$.

In view of the stochastic nature, we assume that $m^{(t)}_d$ is an observed value of some random variable $M_d^{(t)}$.
Since $M^{(t)}_d$ is generated by a maxima operation, standard extreme value theory \citep{Col01} suggests to assume that $M_{d}^{(t)}$ follows the generalized extreme value (GEV) distribution, i.e., 
\[
M_d^{(t)}  \sim \mathrm{GEV}(\mu_d(t), \sigma_d(t), \gamma_d(t))
\]
for some $\mu_d(t), \sigma_d(t)>0, \gamma_d(t) \in \R$,
where the $\mathrm{GEV}(\mu, \sigma, \gamma)$ distribution with location parameter $\mu>0$, scale parameter $\sigma>0$ and shape parameter $\gamma \in \R$ is defined by  its cumulative distribution function
\begin{align}
\label{eq:gevcdf}
G_{(\mu, \sigma, \gamma)}(x) = \exp \Big\{ - \Big(1+\gamma \frac{x-\mu}\sigma\Big)^{-1/\gamma}\Big\}
\end{align}
for $x$ such that $1+\gamma\frac{x-\mu}\sigma>0$.
Due to climate change, the temporal dynamics at location $d$, which are primarily driven by the function $t \mapsto (\mu_d(t), \sigma_d(t), \gamma_d(t))$, are typically non-constant. Any proxy for climate change qualifies as a suitable temporal covariate, and a standard assumption in extreme event attribution studies, motivated by the Clausius–Clapeyron relation, postulates that
\begin{align} \label{eq:ms0}
	\mu_d(t) = \mu_d \exp\left(\frac{\alpha_d \gmst(t)}{ \mu_d} \right), 
	\quad
	\sigma_d(t) = \sigma_d \exp\left(\frac{\alpha_d \gmst(t)}{ \mu_d} \right), 
	\quad
	\gamma_d(t) =\gamma_d
\end{align} 
for certain parameters 
$\alpha_d, \gamma_d \in \R, \mu_d, \sigma_d>0$.
Here, $\gmst(t)$ denotes the smoothed global mean surface temperature anomaly, see \cite{Phi20}. Note that \eqref{eq:ms0} implies
\[
\mathrm{GEV}(\mu_d(t), \sigma_d(t), \gamma_d(t))=
 \exp\left(\frac{\alpha_d \gmst(t)}{ \mu_d} \right) \mathrm{GEV}(\mu_d, \sigma_d, \gamma_d),
\]
hence the model may be identified as a temporal scaling model.  It is further assumed that any temporal dependence at location $d$ is completely due to $\gmst(t)$, which we treat as deterministic and which implies that $M_d^{(1)}, \dots, M_d^{(n)}$ are stochastically independent, for each $d=1, \dots, D$. For the moment, the spatial dependence will be left unspecified.

Recall that the location of interest is the one with $d=1$, which is characterised by the four parameters $\mu_1, \sigma_1, \gamma_1, \alpha_1$. As described before, estimating those parameters based on the observations from location $d=1$ only may be unpleasantly inaccurate, which is why one commonly assumes that the $D$ locations have been carefully selected by experts to meet the following space-time homogeneity assumption:
\begin{align} \label{eq:h0}
\HO: \quad \exists\, \bm \vartheta \in \Theta\  \forall\, d \in \{1, \dots, D\}: \quad \bm \vartheta_d = \bm \vartheta,
\end{align}
where $\Theta:=(0,\infty)^2 \times \R^2$ and $\bm \vartheta=(\mu, \sigma, \gamma, \alpha)^\top, \bm \vartheta_d=(\mu_d, \sigma_d, \gamma_d, \alpha_d)^\top$, and
where the upper index $\mathrm{ED}$ stands for `equal distribution', since, in short, Equation \eqref{eq:h0} states that the location-wise GEV parameters coincide for the $D$ locations.

In the subsequent sections, we aim at testing the validity of the expert's hypothesis $\HO$. Here, it is not only of interest to test the hypothesis for the whole set $\{1, \ldots, D\}$, but also 
to find a (maximal)
subset $A \subset \{1, \ldots, D\}$ with $1 \in A$ and $\abs{A} = k \geq 2$ on which the space-time homogeneity assumption holds. 
Here, for an arbitrary index set $A$, the latter assumption may be expressed through
\begin{align} \label{eq:h0A}
\HO(A): \quad \exists\, \bm \vartheta_A \in \Theta\  \forall\, d \in A: \quad \bm \vartheta_d = \bm \vartheta_A,
\end{align}
with $\Theta$ as in Equation \eqref{eq:h0} and $\bm \vartheta_A=(\mu_A, \sigma_A, \gamma_A, \alpha_A)^\top$, meaning that the location-wise GEV parameters coincide for all locations with index in the set $A$, making the respective locations a possible pooling region. 

Now, a maximal subset $A$ for which Equation \eqref{eq:h0A} holds may be determined with the following strategy:
Since we are interested in finding all locations that `match' the location of primary interest with index $ d=1$, we test for each pair $A_d = \{1, d\}, d = 2, \ldots, D$, whether the null hypothesis $\HO(A_d)$ holds. This will provide us with a set of p-values based on which we can decide which locations to reject and which not to reject. 
Those locations that are not rejected can then be assumed to be sufficiently homogeneous and are thus included in the suggestion of a pooling region of maximal extent. For further details on this strategy and the impact of the induced multiple testing problem, see Section \ref{subsec:combineTeststats}.

\subsection{Coordinate-wise maximum likelihood estimation} 
\label{subsec:ml}

The starting point for the subsequent test statistics are the coordinate-wise maximum likelihood estimators for the model specified in \eqref{eq:ms0}. Writing $c^{(t)}=\gmst(t)$ for brevity, the log-likelihood contribution of observation $(M_d^{(t)}, c^{(t)})$ is given by $\ell_{\bm \vartheta_d}(M_d^{(t)}, c^{(t)})$, where
\begin{align} \label{eq:ld}
\ell_{\bm \vartheta_d}(x,c) = \log g_{(\mu_d \exp(\alpha_d c/\mu_d), \sigma_d \exp(\alpha_d c/\mu_d), \gamma_d)}(x) 
\end{align}
with $g_{(\gamma, \mu, \sigma)}(x) = \frac\partial{\partial x} G_{(\mu, \sigma, \gamma)}(x)$ the probability density function of the $\mathrm{GEV}(\mu, \sigma, \gamma)$-distribution.
The maximum likelihood estimator for $\bm{\vartheta}_d$ at location $d$ is then defined as
\begin{align} \label{eq:ml}
\hat{\bm \vartheta}_d  \in \argmax_{\bm \vartheta_d \in \Theta} \sum_{t=1}^n \ell_{\bm \vartheta_d}(M_d^{(t)}, c^{(t)}).
\end{align}
The arg-maximum cannot be calculated explicitly, but may be found by suitable numerical optimization routines.
We denote the gradient and the Hessian matrix of $\bm \vartheta \mapsto \ell_{\bm \vartheta}(x,c)$ by $\dot \ell_{\bm \vartheta}(x,c) \in \R^4$ and  $\ddot \ell_{\bm \vartheta}(x,c) \in \R^{4\times 4}$, respectively. 
Under appropriate regularity assumptions, standard asymptotic expansions (\citealp{Van98}, see also \citealp{BucSeg17} for the stationary GEV family) imply that $\hat {\bm \theta}=(\hat{\bm \vartheta}_1^\top, \dots, \hat{\bm \vartheta}_D^\top)^\top \in \Theta^{D}$ is approximately Gaussian with mean ${\bm \theta}=({\bm \vartheta}_1^\top, \dots, {\bm \vartheta}_D^\top)^\top$ and covariance $n^{-1} \bm \Sigma_n$, where $\bm \Sigma_n = (\bm \Sigma_{n;j,k})_{j,k=1}^D \in \R^{4D \times 4D}$ is defined as
\begin{align} \label{eq:apprSigma}
\bm \Sigma_{n;j,k} &= J_{n,j,\bm{\vartheta}_{j}}^{-1}  \Big( \frac{1}{n}\sum_{t = 1}^n \Cov\big[ \dot\ell_{\bm\vartheta_j}(M_j^{(t)}, \ct),    \dot\ell_{\bm\vartheta_k}(M_k^{(t)}, \ct) \big] \Big) J_{n,k,\bm{\vartheta}_{k}}^{-1}  \in \R^{4 \times 4}
\end{align} 
with 
$
J_{n,j,\bm\vartheta} =  \frac{1}{n}\sum_{t = 1}^n \Exp[ \ddot\ell_{\bm\vartheta}(M_j^{(t)}, \ct) ] \in \R^{4\times 4}$. See Appendix~\ref{app:mlest} for details and Appendix~\ref{app:covest} for a suitable estimator $\hat{\bm \Sigma}_n$ for $\bm \Sigma_n$.

\subsection{Wald-type test statistics}\label{subsec:waldtest}

We define test statistics which allow to test for the sub-hypotheses $\HO(A)$ of $\HO$ from Equation \eqref{eq:h0A}, where $A \subset \{1, \ldots, D\}$.
For that purpose, we propose to use classical Wald-type test statistics; see Section 14.4.2 in \cite{Leh21} for a general discussion and \cite{Lil22} for a similar approach in temporally stationary GEV models, i.e., with $\alpha_d$ fixed to $\alpha_d=0$. 

Write $A=\{d_1, \dots, d_k\}$ with $1 \le d_1 < \dots < d_k \le D$ and let $h_A: \R^{4D} \to \R^{4(k-1)}$ be defined by 
\begin{align*}
h_A(\bm \theta) = h_A(\bm\vartheta_{1}, \ldots, \bm\vartheta_{D}) 
& = 
(\bm\vartheta_{d_1}^\top - \bm\vartheta_{d_2}^\top, \bm\vartheta_{d_2}^\top - \bm\vartheta_{d_3}^\top, \ldots, \bm\vartheta_{d_{k-1}}^\top - \bm\vartheta_{d_k}^\top)^\top \\
& = 
(\mu_{d_1} - \mu_{d_2}, \sigma_{d_1} - \sigma_{d_2}, \gamma_{d_1} - \gamma_{d_2}, \alpha_{d_1} - \alpha_{d_2},  \\ 
& \hspace{4cm }\ldots,  \gamma_{d_{k-1}} - \gamma_{d_k}, \alpha_{d_{k-1}} - \alpha_{d_k})^\top.
\end{align*}
We may then write $\HO(A)$ equivalently as 
\[
\HO(A):  h_A(\bm \theta) = 0.
\]
Hence, significant deviations of $h_A(\hat{\bm \theta})$ from 0 with $\hat{\bm \theta}$ from Section~\ref{subsec:ml} provide evidence against $\HO(A)$. Such deviations may be measured by the Wald-type test statistic
\begin{equation}\label{eq:waldteststatistic}
T_n(A) = n\,
(h_A(\hat{\bm\theta}))^\top \, 
\left(\bm H_A \;\hat{\bm \Sigma}_n\;\bm H_A^\top\right)^{-1} \, 
h_A(\hat{\bm\theta}), 
\end{equation}
where $\bm H_A= \dot h_A(\bm \theta) \in \R^{4(k-1) \times 4D}$ denotes the Jacobian matrix of $\bm \theta \mapsto h_A(\bm \theta)$, which is a matrix with entries in $\{-1,0,1\}$ that does not depend on $\bm \theta$. In view of the asymptotic normality of $\hat{\bm\theta}$, see Section~\ref{subsec:ml}, the asymptotic distribution of $T_n(A)$ under the null hypothesis $\HO(A)$ is the chi-square distribution $\chi_{4(k-1)}^2$ with $4(k-1)$ degrees of freedom; see also Section 4 in \cite{Lil22}. Hence, rejecting $\HO(A)$ if $T_n(A)$ exceeds the $(1-\alpha)$-quantile of the  $\chi_{4(k-1)}^2$-distribution provides a statistical test of asymptotic level $\alpha \in (0,1)$. 
The finite-sample performance of the related test in the stationary setting was found to be quite inaccurate (see \citealp{Lil22}). To overcome this issue, we propose a suitable bootstrap scheme in the next section.

\subsection{Parametric bootstrap devices for deriving p-values}\label{sec:boot}

Throughout this section, we propose two bootstrap devices that allow to simulate approximate samples from the $\HO(A)$-distribution of the test statistic $T_n(A)$ from Equation \eqref{eq:waldteststatistic}. 
Based on a suitably large set of such samples, one can compute a reliable p-value for testing $\HO(A)$, even for short sample sizes. 

The first method is based on a global fit of a max-stable process model to the entire region under consideration, while the second one is based on fitting multiple pairwise models. The main difference of the two approaches is that the first one can test the hypothesis $\HO(A)$ for arbitrary subsets $A \subset \{1, \ldots, D\},$ while the second approach is restricted to testing the null hypothesis on subsets of cardinality two, i.e., it can only test whether a pair of locations is homogeneous. Depending on the question that is asked, applying the one or the other method may be advantageous.

\subsubsection{Global bootstrap based on max-stable process models}\label{subsec:boot_maxstable} 

The subsequent bootstrap device is a modification of the parametric bootstrap procedure described in Section 5.3 of \cite{Lil22}.
Fix some large number $B$, say $B=200$, noting that larger numbers are typically  better, but going beyond $B=1000$ is usually not worth the extra computational effort.

The basic idea is as follows: for each $b=1, \dots, B$, simulate artificial \textit{bootstrap samples} 
\[
\mathcal D^*_b = \big\{M_{d,b}^{(t),*}: t\in\{1, \dots, n\},d \in \{1, \dots, D\} \big\}
\] 
that have a sufficiently similar spatial dependence structure as the observed data $\mathcal D=\{M_{d}^{(t)}: t\in\{1, \dots, T\},d \in \{1, \dots, D\}\}$ and that satisfy the null hypothesis $\HO$. For each fixed $A\subset \{1, \dots, D\}$ with $k=|A|\ge 2$, the test statistics computed on all bootstrap samples, say $(T_{n,b}^*(A))_{b=1, \dots, B}$,  are then compared to the observed test statistic $T_n(A)$. Since the bootstrap samples do satisfy $\HO(A)$, the observed test statistic $T_n(A)$ should differ significantly from the bootstrapped test statistics in case $\HO(A)$ is not satisfied on the observed data.

Here, for simulating the bootstrap samples, we assume that the spatial dependence structure of the observed data can be sufficiently captured by a max-stable process model. Max-stable processes provide a natural choice here, since they are the only processes that can arise, after proper affine transformation, as the limit of maxima of independent and identically distributed random fields $\{Y_i(x) : x \in \R^p\}$ (\citealp{Col01}, Section 9.3).
Parametric models for max-stable processes are usually stated for unit Fréchet (i.e., $\GEV(1,1,1)$) margins. Therefore, the first steps in our algorithm below aim at transforming the margins of our observed data to be approximately unit Fréchet. 

More precisely, the parametric bootstrap algorithm is defined as follows:

\begin{algorithm}[Bootstrap based on max-stable processes] \label{alg1}~
\begin{compactenum}[(1)]
    \item For each $d  \in\{1, \ldots, D\}$, calculate $\hat{\bm \vartheta}_d$ from Section~\ref{subsec:ml}.
    \item For each $d  \in\{1, \ldots, D\}$, transform the observations to approximately i.i.d.\ Fréchet-distributed data, by letting 
    \begin{align}\label{eq:gev2frech}
    Y_d^{(t)} =\left\{ 1 + \hat \gamma_d \frac{M_d^{(t)} - \hat \mu_d \exp\left(\frac{\hat \alpha_d \gmst(t)}{ \hat \mu_d} \right)}{\hat \sigma_d \exp\left(\frac{\hat \alpha_d \gmst(t)}{ \hat \mu_d} \right)} \right\}_+^{1 /\gamma_d} \quad  (t\in\{1, \dots, n\}).
    \end{align}
    \item Fit a set of candidate max-stable process models with standard Fréchet margins  to the observations $(Y_1^{(t)}, \ldots, Y_D^{(t)})_{t=1, \dots, n}$ and choose the best fit according to the composite likelihood information criterion (CLIC), which is a model selection criterion that is commonly applied when fitting max-stable process models. 
   Throughout, we chose the following three models:
    \begin{compactenum}
        \item  Smith's model (3 parameters);
        \item Schlather's model with a powered exponential correlation function (3 parameters);
        \item the Brown-Resnick process (2 parameters).
    \end{compactenum}
    For further details on max-stable processes, the mentioned models and the CLIC, see \cite{davisonModelSpatExt} and \cite{Dav12}. Respective functions are implemented in the \texttt{R} package \texttt{SpatialExtremes} \citep{SpatialExtremes}.
    \item For $ b \in\{ 1, \ldots, B\}$ and $t \in\{1, \dots, n\}$, simulate spatial data with unit Fréchet margins from the chosen max-stable process model, denoted by
    \[ 
    (Y_{1,b}^{(t), \ast}, Y_{2,b}^{(t), \ast}, \ldots, Y_{D,b}^{(t), \ast}).
    \]
    \end{compactenum}
    
\medskip
\noindent Note that until now we haven't used the particular hypothesis $\HO(A)$. Subsequently, fix $A=\{d_1, \dots, d_k\}$ with $1 \le d_1< \dots < d_k \le D$.
\medskip

\begin{compactenum}[(1)]   
\item[(5)] Assume that $\HO(A)$ from Equation \eqref{eq:h0A} is true, and estimate the four dimensional model parameters $\bm \vartheta_A = (\mu_A,\sigma_A,\gamma_A, \alpha_A)^\top \in \Theta$ by (pseudo) maximum likelihood based on the pooled sample 
    \begin{align*}
     (M_{d_1}^{(1)}, c^{(1)}), \dots, (M_{d_1}^{(n)}, c^{(n)}), (M_{d_2}^{(1)}, c^{(1)}), \dots, (&M_{d_2}^{(n)}, c^{(n)}), \dots \\
     \dots, (&M_{d_k}^{(1)}, c^{(1)}), \dots, (M_{d_k}^{(n)}, c^{(n)}).
    \end{align*}
    Denote the resulting parameter vector as $\hat{\bm\vartheta}_{A}  = (\hat\mu_A, \hat\sigma_A, \hat\gamma_A, \hat\alpha_A)^\top$, and note that $\hat{\bm\vartheta}_{A}$ should be close to $\hat{\bm\vartheta}_{d}$ for each $d\in A$, if $\HO(A)$ is met.
\item[(6)] Transform the margins of the bootstrap samples to the ones of a GEV-model satisfying $\HO(A)$, by letting
    \begin{align}\label{eq:frech2gev}
    M_{d,b}^{(t), \ast} =\hat \mu_A \exp\left(\frac{\hat \alpha_A \gmst(t)}{ \hat \mu_A} \right)
    + 
    \hat \sigma_A \exp\left(\frac{\hat \alpha_A \gmst(t)}{ \hat \mu_A} \right) \frac{(Y_{d,b}^{(t), \ast})^{\hat\gamma_A} -1}{\hat\gamma_A} 
    \end{align}
    for $ t\in\{1, \ldots, n\}, d \in A$ and $b \in\{ 1, \ldots, B\}$. For each resulting bootstrap sample $\mathcal D_b^*(A) =  \{ M_{d,b}^{(t), \ast }: t \in \{1, \dots, n\}, d \in A\}$, compute the value $t_{n,b}^\ast(A)$ of the test statistic $T_n(A)$ from Equation \eqref{eq:waldteststatistic}. Note that $T_n(A)$ only depends on the coordinates with $d \in A$.
\item[(7)] Compute the value $t_n(A)$ of the test statistic $T_n(A)$ from Equation \eqref{eq:waldteststatistic} on the observed sample.
\item[(8)]
Compute the bootstrapped $p$-value by
\begin{equation*}
    p(A) = \frac{1}{B+1} \sum_{b=1}^{B} \ind(t_n(A) \leq t_{n,b}^\ast(A)).
\end{equation*}        
\end{compactenum}
\end{algorithm}

In a classical test situation, one may now reject $\HO(A)$ for a fixed set $A$ at significance level $\alpha \in (0,1)$ if $p(A) \le \alpha$. In the current pooling situation, we would need to apply the test to multiple pooling regions $A$, which hence constitutes a multiple testing problem where standard approaches yield inflated levels. We discuss possible remedies in Section~\ref{subsec:combineTeststats}.

\subsubsection{Pairwise bootstrap based on bivariate extreme value distributions}

Recall that the location of primary interest is the one with index $d=1$. 

As stated in Section \ref{subsec:model}, it is of interest to test for all bivariate hypotheses $\HO(\{1, d\})$ with $d=2, \dots, D$. For that purpose, we may apply a modification of the bootstrap procedure from the previous section that makes use of bivariate extreme value models only.  By doing so, we decrease the model risk implied by imposing a possibly restrictive global max stable process model. 

The modification only affects step (3) and (4) from Algorithm \ref{alg1}. More precisely, for testing the hypothesis $\HO(A_d)$ with $A_d=\{1, d\}$ for some fixed value $d=2, \dots, D$, we make the following modifications:

\begin{algorithm}[Pairwise bootstrap based on bivariate extreme value distributions] \label{alg2}

\noindent
Perform step (1) and (2) from Algorithm \ref{alg1} with the set $\{1, \ldots, D\}$ replaced by $A_d$.
\begin{compactenum}
    \item[(3a)] Fit a set of bivariate extreme value distributions to the bivariate sample $(Y_1^{(t)}, Y_d^{(t)})_{t=1, \dots , n}$, assuming the marginal distributions to be unit Fréchet. Choose the best fit according to the Akaike information criterion (AIC), a model selection criterion that rewards a good fit of a model and penalises the model's complexity at the same time \citep{Aka73}.
    Possible models are: 
        \begin{compactenum}
        \item the Hüsler-Reiss model (1 parameter);
        \item the logistic model (1 parameter); 
        \item the asymmetric logistic model (2 parameters).
    \end{compactenum}
    Note that all models are implemented in  \cite{evd}.
    \item[(4a)] For $b \in \{1, \ldots, B\}$  and $t \in\{1, \dots, n\}$, simulate bivariate data with unit Fréchet margins from the chosen bivariate extreme value model, denoted by
    $
    (Y_{1,b}^{(t), \ast}, Y_{d,b}^{(t), \ast}).
    $
\end{compactenum}
Perform Steps (5)-(8) from Algorithm \ref{alg1} with $A=A_d$.
\end{algorithm}

Note that Algorithm~\ref{alg2} is computationally more expensive than Algorithm~\ref{alg1} since model selection and fitting of dependence models and its subsequent simulation must be performed separately for each hypothesis $\HO(A_d)$ of interest.

\subsection{Combining test statistics}\label{subsec:combineTeststats}

As already addressed at the end of Section \ref{subsec:model}, it is not only of interest to test the global hypothesis $\HO$, since a possible rejection of $\HO$ gives no indication about which locations deviate from the one of primary interest. Instead, one might want to test hypotheses on several subsets and then pool those subsets for which no signal of heterogeneity was found. 
In this subsection, we provide the mathematical framework of testing sub-hypotheses and discuss how to deal with the induced multiple testing problem.

Mathematically, we propose to regard $\HO$ as a global hypothesis that is built up from elementary hypotheses of smaller dimension. A particularly useful decomposition is based on pairwise elementary hypotheses: recalling the notation $\HO(A)$ from Equation \eqref{eq:h0A}, we clearly have
\begin{align}\label{eq:elemHypSchnitt}
\HO 
=
\bigcap_{d=2}^D \HO(\{1,d\}),
\end{align}
i.e., $\HO$ holds globally when it holds locally for all pairs $\{1, d\}$ with $d \in\{2, \ldots, D\}.$
We may now either apply Algorithm \ref{alg1} or Algorithm \ref{alg2} to obtain a p-value, say $p^{\text{raw}}_d=p(\{1,d\})$, for testing $\HO(\{1,d\})$, for any $d\in\{2, \dots, D\}$. Each p-value may be interpreted as a signal for heterogeneity between locations $1$ and $d$, with smaller values indicating stronger heterogeneity. The obtained raw list of p-values may hence be regarded as an exploratory tool for identifying possible heterogeneities.

Since we are now dealing with a multiple testing problem, it might be advisable to adjust for multiple comparison in order to control error rates. 
This can be done by 
interpreting the raw list based on classical statistical testing routines, in which p-values are compared with suitable critical values to declare a hypothesis significant. Several methods appear to be meaningful, and we discuss three of them in the following.
For this, let $\alpha\in(0,1)$ denote a significance level, e.g., $\alpha=0.1$.

\medskip
\noindent
\textbf{IM (Ignore multiplicity):} reject homogeneity for all pairs $\{1,d\}$ for which $p_d^{\text{raw}}\le \alpha$. In doing so, we do not have any control over false rejections. In particular, in case $D$ is large, false rejections of some null hypotheses will be very likely. On the other hand, the procedure will have decent power properties, and will likely detect most alternatives. Hence, in a subsequent analysis based on the pooled sample of homogeneous locations, we can expect estimators to exhibit comparably little bias and large variance.

\medskip
\noindent
\textbf{Holm (Control the family-wise error rate):} 
apply Holm's stepdown procedure (\citealp{Hol79}). For that purpose, sort the p-values $p_j=p^{\text{raw}}_{1+j}=p(\{1,1+j\})$ with $j=1, \dots, D-1$; denote them by $p_{(1)} \le \dots \le p_{(D-1)}$. Starting from $j=1$, determine the smallest index $j$ such that 
\[
p_{(j)} > \alpha_j := \alpha/(D-j).
\] 
If $j=1$, then reject no hypotheses. If no such index exists, then reject all hypotheses. Otherwise, if $j \in \{2, \dots, D-1\}$, reject the hypotheses that belong to the p-values $p_{(1)}, \dots, p_{(j-1)}$. 

The procedure can be equivalently expressed by adjusted p-values. Recursively defining
$ \tilde{p}_{(1)} = \min\{1,  (D-1)p_{(1)} \} $ and
\[ 
\tilde{p}_{(j)} = \min\left\{1,  \max\{ \tilde p_{(j-1)}, (D -j) p_{(j)} \} \right\}
\]
for $j = 2, \ldots, D-1$, we simply reject those hypotheses that belong to the adjusted p-values with $\tilde p_{(j)} \le \alpha$.

Holm's stepdown procedure is known to asymptotically control the family-wise error rate (FWER) at level $\alpha$, i.e., 
\[
\FWER := \Pr\big( \text{reject any true null hypothesis } \HO(\{1,d\}) \big) \le \alpha,
\]
see Theorem 9.1.2 in  \cite{Leh21}.

In general, controlling the family-wise error rate will result in comparably little power, i.e., we might falsely identify some pairs of locations as homogeneous. Hence, in a subsequent analysis based on the pooled sample of homogeneous locations, we can expect estimators to exhibit comparably large bias and little variance.

\medskip
\noindent
\textbf{BH (Control the false discovery rate):} apply the Benjamini Hochberg stepup procedure 
\citep{Ben95}. For that purpose, sort the p-values $p_j=p^{\text{raw}}_{1+j}=p(\{1,1+j\})$ with $j=1, \dots, D-1$; denote them by $p_{(1)} \le \dots \le p_{(D-1)}$. Starting from $j=D-1$, determine the largest index $j$ such that 
\[
p_{(j)}  \le  \alpha_j := \frac{j \alpha}{ (D-1)}.
\]
If no such index exists, then reject no hypotheses. Otherwise, if $j \in \{1, \dots, D-1\}$, reject the hypotheses that belong to the p-values $p_{(1)}, \dots, p_{(j)}$. 

Again, one can compute adjusted $p$-values $\tilde p_{(j)}$ such that the procedure is equivalent to rejecting those hypotheses for which $\tilde p_{(j)} \le \alpha$. For that purpose, let
$\tilde{p}_{(D-1)} = \min\{1, (D-1) p_{(D-1)} \}$ and recursively define, for $j = D-2, \ldots, 1$,
\[ 
\tilde{p}_{(j)} = \min\left\{ 1, \min\left\{ (D-1) \frac{p_{(j)}}{j}, \tilde{p}_{(j+1)} \right\} \right\}. 
\]

Under an additional assumption on the p-values that belong to the true null hypotheses (they must exhibit some positive dependence), the BH procedure is known to asymptotically control the false discovery rate (FDR) at level $\alpha$, i.e., 
\[
\FDR := \mathbb E\Big[ \frac{\text{Number of false rejections}}{\text{Number of all rejections}} \bm 1(\text{at least one rejections}) \Big] \le \alpha,
\]
see Theorem 9.3.3 in \cite{Leh21}. Control of the FDR will be confirmed by the simulation experiments in Section~\ref{sec:sim}. 

If one were interested in guaranteed theoretical control of the FDR rate, one might alternatively apply the Benjamini Yekutieli (BY) stepup procedure, see \citep{Ben01} and Theorem 9.3.3 in  \cite{Leh21}. In view of the fact that the procedure is much more conservative than BH, we do not recommend its application in the current setting.

Concerning a subsequent analysis, estimators based on a pooled sample obtained from the BH procedure can be expected to exhibit bias and variance to be somewhere between the IM and Holm procedure.

\begin{remark} \label{rem:snake}
The decomposition of $\HO$ into hypotheses of smaller dimensionality is not unique. For instance, we may alternatively write
\begin{align}\label{eq:elemHypSchnitt2}
\HO 
=
\bigcap_{k=1}^K \HO( B_k),
\end{align}
where
$\{1 \} \subset B_1 \subset B_2 \dots \subset B_K = \{1, \dots,d \}$ denotes an increasing sequence of regions with $2 \le |B_1| < |B_2| < \dots < |B_K|=d$ (for instance, $B_k=\{1, 2,\dots, 1+k\}$ with $k=1, \dots, D-1$). In practice, the sequence is supposed to be derived from some expert knowledge of the region of interest; it shall represent a sequence of possible pooling regions where $B_k$ is constructed from $B_{k-1}$ by adding the locations which are a priori `most likely' homogeneous to the locations in $B_k$. Note that, necessarily, $K \le D-1$, which provides an upper bound on the number of hypotheses to be tested. 

The derivation of respective testing methods is straightforward. In view of the facts that the choice of the sequence is fairly subjective and that the eventual results crucially depend on that choice, we do not pursue the method any further. 
\end{remark}

\section{Simulation Study} 
\label{sec:sim}

A large-scale Monte Carlo simulation study was conducted to assess the performance of the proposed bootstrap procedures in finite sample situations. 
We aim at answering the following questions:
\begin{compactenum}[(a)]
    \item Regarding the test's \textit{power}: What percentage of locations that are heterogeneous w.r.t.\ the location of primary interest can be expected to be identified correctly?
    \item Regarding the test's \textit{error rates}: What percentage of locations that are homogeneous w.r.t.\ the location of primary interest can be expected to be wrongly identified as heterogeneous (FDR)? What is the probability of wrongly identifying at least one location that is homogeneous w.r.t.\ the location of interest as heterogeneous (FWER)?
    \item Regarding the chosen pooling regions: How does return level (RL) estimation based on the pooling regions proposed by the bootstrap procedures compare to RL estimation based on the location of interest only or the whole (heterogeneous) region? 
\end{compactenum}

\begin{figure}[t]
    \centering
    \begin{minipage}{.5\textwidth}
        \centering
        \includegraphics[width=0.9\textwidth]{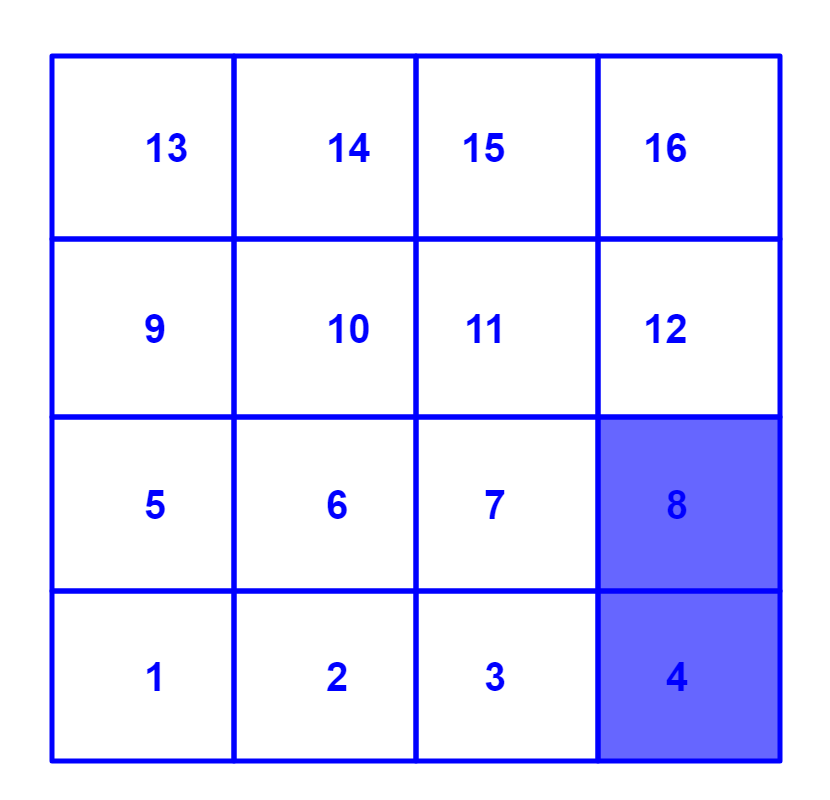}
    \end{minipage}%
    \begin{minipage}{0.5\textwidth}
        \centering
        \includegraphics[width=0.9\textwidth]{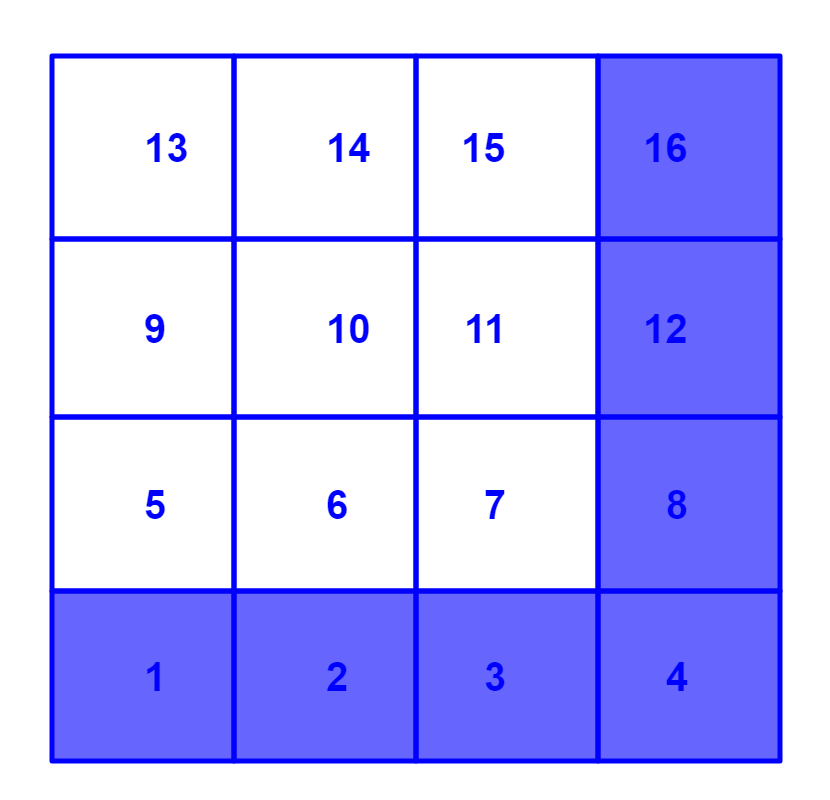}
    \end{minipage}
    \caption{Illustration of the grid used for the simulation. The regions contained in $A_{\text{dev}}$ are shaded in blue, with $|A_{\text{dev}} | = 2$ shown in the left plot and $|A_{\text{dev}} | = 7$ shown on the right. The region of interest is the one labelled 10.}
        \label{fig:sim_design}
\end{figure}

The data was generated in such a way that the temporal spatial dynamics from the case study in Section \ref{sec:case} are mimicked.
To achieve this, we started by fitting the scale-GEV model from Equation \eqref{eq:ms0} to annual block-maxima of observations from 1950--2021 at 16 spatial locations  in Western Europe (i.e., $n=72$ and $D=16$) that are arranged in a $4\times4$ grid; see Figure~\ref{fig:sim_design} and the additional explanations in Section~\ref{sec:case}. 
The locations correspond to the center points of the grid cells; the distance between the center points of two neighbouring grid cells is approximately 140 km.
The location-wise GEV parameter estimates $\hat{\bm \vartheta}_d$ exhibit the following approximate ranges over $d\in\{1, \dots, 16\}$: $\hat\mu_d \in (18.1, 30.8)$ with a mean of $20.85$, $\hat\sigma_d \in (4.185, 7.92)$ with a mean of 5.3, $\hat\gamma_d \in (-0.13, 0.36)$ with a mean of 0.08 and $\hat\alpha_d \in (-2.3, 5.08) $ with a mean of 1.5. Fitting the scale-GEV model to the full pooled sample of size $n\cdot D=1152$, we obtained parameter estimates that were close to the means over the location-wise parameter estimates, with $20.37, 5.8, 0.1, 1.5$ for location, scale, shape and trend parameter, respectively.
Next, we transformed the margins to (approximate) unit Fréchet by applying the transformation from Equation \eqref{eq:gev2frech}, such that we can fit several max-stable process models to the transformed data. The best fit was Smith's model with approximate dependence parameters $\sigma_{11} = 0.4, \sigma_{12} = 0.2, \sigma_{22} = 0.9$; see \cite{davisonModelSpatExt} for details on the model.

Based on these model fits,
we chose to generate data with the following specifications: first, the sample size was fixed to $n=75$ and the regional $4\times 4$ grid was chosen as described before, i.e., $d=16$. The grid cell/location labelled `$10$' is chosen as the one of primary interest. 
Further, the dependence structure is fixed to that of Smith's model with (approximately) those parameters that gave the best fit on the observed data, i.e.  $\sigma_{11} = 0.4, \sigma_{12} = 0.2, \sigma_{22} = 0.9$.
For simulating data, we first simulate from this max-stable process model \citep{SpatialExtremes} and then transform the margins to scale-GEV distributions, either in a homogeneous or in a heterogeneous manner.
Here, the globally homogeneous model is defined by fixing the marginal scale-GEV parameters to
approximately the mean values of the location-wise GEV parameters obtained for the real observations, i.e.,
\begin{align} \label{eq:sim0}
\mu_{d} = 20, \ \sigma_{d} = 5.5, \ \gamma_d = 0.1, \ \alpha_d = 1.5
\end{align}
for each $d\in\{1, \dots, 16\}$.

Starting from this homogeneous model, we consider two different heterogeneous scenarios. In the first scenario, we fix $\bm \vartheta_d=(\mu_d, \sigma_d, \gamma_d, \alpha_d)^\top$ as in Equation \eqref{eq:sim0} for all $d \in A_{\text{hom}} = \{1, \dots, 16\}\setminus \{ 4,8\}$, while 
\begin{gather} \label{eq:altmod}
 \begin{aligned}
  \mu_{d} &= 20  + c_\mu, & \hspace{.5cm} &  c_\mu \in \{ -3, -1.5, 0, 1.5, 3\} , \\ 
  \sigma_{d} &= 5.5 \cdot c_\sigma,  & \hspace{.5cm} &  c_\sigma \in \{ 0.7, 0.85, 1, 1.15, 1.3\}, \\ 
  \gamma_d &= 0.1 + c_\gamma, & \hspace{.5cm} & c_\gamma \in \{ -0.1, 0, 0.1\} ,\\
  \alpha_d &= 1.5 + c_\alpha,& \hspace{.5cm} & c_\alpha \in \{-1,0,1 \},
  \end{aligned}
 \end{gather}
for $d\in A_{\text{dev}} = \{4,8\}$ with $(c_\mu, c_\sigma, c_\gamma, c_\alpha)\ne(0,0,0,0)$. Note that this defines $5 \cdot 5 \cdot 3 \cdot 3 - 1 = 224$ different heterogeneous models. In the second scenario, we consider the same construction with $A_{\text{hom}}=\{5, 6, 7, 9, 10, 11, 13, 14, 15\}$ and $A_{\text{dev}} = \{1,2,3,4,8,12,16\}$. 
An illustration of the grid cells and their partition into homogeneous and non-homogeneous areas can be found in Figure  \ref{fig:sim_design}.
Overall, we obtain 448 different heterogeneous models and one homogeneous model.

For each of the 449 models, we now apply the following three bootstrap procedures, each carried out with $B = 300$ bootstrap replications (recall that the grid cell of interest is the one labelled with 10):
\begin{compactenum}[(B1)]
\renewcommand{\theenumi}{(B1)}
\renewcommand{\labelenumi}{\theenumi}
    \item\label{item:b1} The bootstrap procedure from Algorithm~\ref{alg1} with $A=\{1, \dots, 16\}$.
\renewcommand{\theenumi}{(B2)}
\renewcommand{\labelenumi}{\theenumi}
    \item\label{item:b2}  The bootstrap procedure from Algorithm~\ref{alg1} for all sets $A_d=\{10,d\}$ with $d \in \{1, \ldots, 16\}\setminus \{10\} $.
\renewcommand{\theenumi}{(B3)}
\renewcommand{\labelenumi}{\theenumi}
    \item\label{item:b3} The bootstrap procedure from Algorithm~\ref{alg2} for all sets $A_d=\{10,d\}$ with $d \in \{1, \ldots, 16\}\setminus \{10\}$.
\end{compactenum}
Note that the second and third method both yield 15 raw p-values. Each procedure was applied to 500 simulated samples from all models under consideration. 

\begin{figure}[t!]
\makebox[\textwidth][c]{
 	\includegraphics[width=1.2\textwidth]{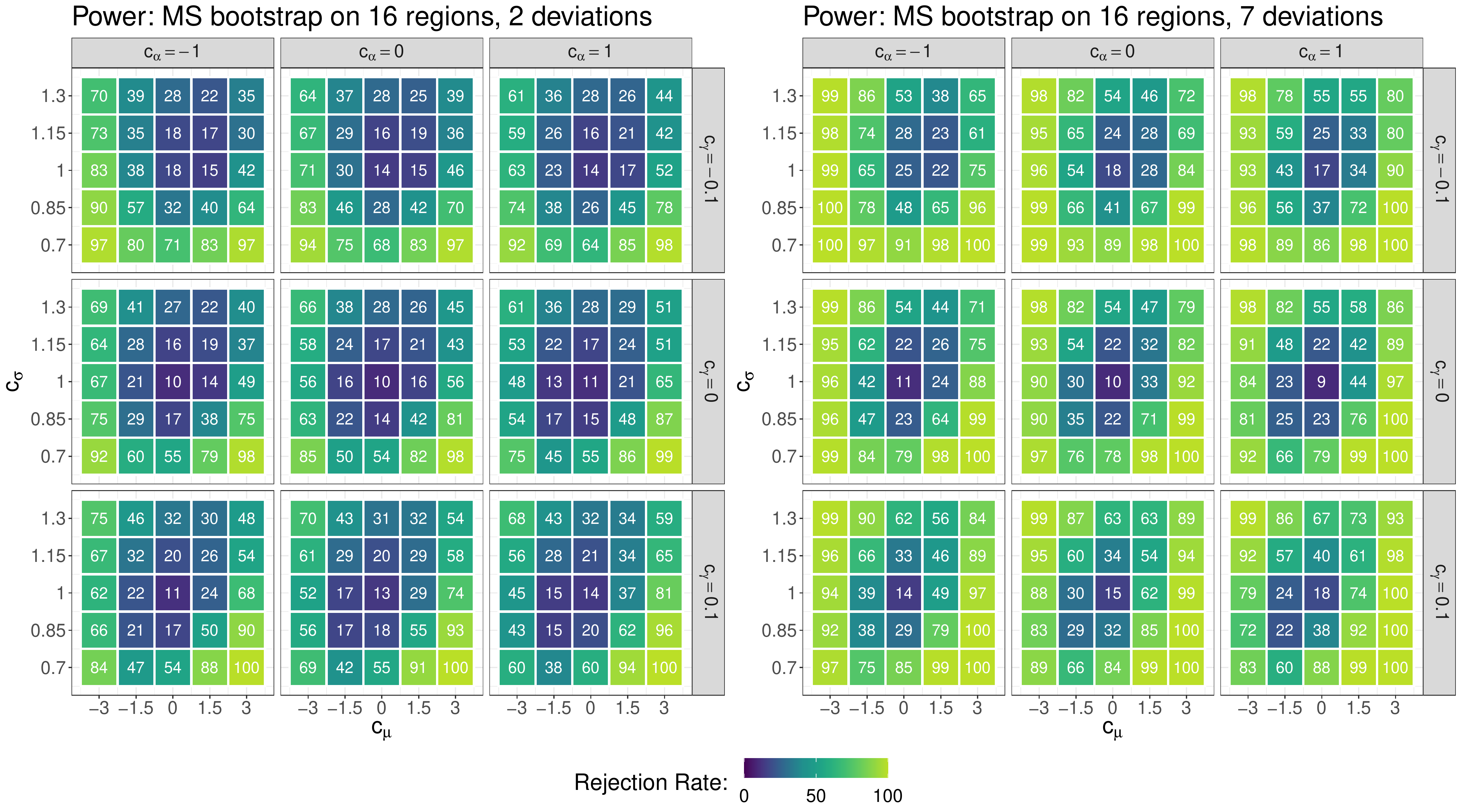}
}
\caption{\label{fig:power_ms_on16} Rejection rates in $\%$ obtained for \ref{item:b1}, in the setting where either 2 (left plot) or 7 (right plot) regions deviate from the others. 
Each coloured square contains the rejection rate for one of the 225 different models, with the central square with $c_\mu=c_\sigma=0$ corresponding to the null hypothesis.
The $x$- and $y$-axis and the facets determine the values of the scale-GEV parameter vector of the deviating locations through Equation \eqref{eq:altmod}.
}
\end{figure}

Regarding \ref{item:b1}, we compute the percentage of rejections among the 500 replications, which represents the empirical type \uproman{1} error of the test under the homogeneous model and the empirical power under the heterogeneous models. The results can be found in Figure \ref{fig:power_ms_on16}. The null hypothesis is met in the central square only, and we observe that the nominal level of $\alpha=0.1$ is perfectly matched. All non-central squares correspond to different alternatives, and we observe decent power properties in both scenarios. Note that a rejection only implies that the entire region $\{1, \dots, 16\}$ is not homogeneous; there is no information on possible smaller subgroups that are homogeneous to the location of interest.

Regarding \ref{item:b2} and \ref{item:b3}, rejection decisions were obtained for each hypothesis $\HO(\{10,d\})$ by one of the three methods from Section~\ref{subsec:combineTeststats}. The empirical family-wise error rate is then the percentage of cases (over 500 replications) for which at least one null hypothesis was rejected. Likewise, for the false discovery rate, we calculate, for each replication, the number of false rejections and divide that by the total number of rejections (when the number of total rejections is 0, this ratio is set to 0). The empirical false discovery rate is obtained by taking the mean over all 500 replications.
Similarly, for assessing the power properties, we calculate the empirical proportion of correct rejections (i.e., among the 2 or 7 locations that deviate, the proportion of detected heterogeneous locations) over all 500 replications.

Results for the false discovery and family-wise error rate are given in Table~\ref{tab:fdr}. We find that the $p$-value combination methods from Section \ref{subsec:combineTeststats} are sufficiently accurate: the BH method controls the false discovery rate, while Holm's method controls the family-wise error rate. This holds exactly for procedures \ref{item:b3}, where the maximal FDR (FWER) of the BH (Holm) method is at 9.4\% (8.7\%), and approximately for \ref{item:b2}, where the maximal FDR (FWER) is at 12.2\% (12.6\%).
Further, we see that the IM procedure neither controls the FWER nor the FDR.

\begin{table}[t]
\centering
\makebox[\textwidth][c]{
\footnotesize
\begin{tabular}{l|| rr|rr|rr || rr|rr|rr}
\hline 
Method & \multicolumn{2}{c|}{min FDR} & \multicolumn{2}{c|}{max FDR} & \multicolumn{2}{c||}{mean FDR} & \multicolumn{2}{c|}{min FWER} & \multicolumn{2}{c|}{max FWER} & \multicolumn{2}{c}{mean FWER}  \\ 
& (B2) & (B3) & (B2) & (B3) & (B2) & (B3) & (B2) & (B3) & (B2) & (B3) & (B2) & (B3)  \\ 
  \hline \hline \addlinespace[.2cm]
  \multicolumn{13}{l}{\quad\textit{Scenario 1: $|A_\dev|=2$}} \\ \hline
BH & 7.3 & 5.6 & 12.2 & 9.4 & 9.4 & 7.5 & 9.1 & 6.9 & 21.2 & 19.0 & 14.4 & 11.8 \\ 
  Holm & 3.0 & 2.3 & 11.7 & 8.3 & 7.1 & 5.1 & 6.9 & 5.0 & 12.6 & 8.7 & 9.5 & 6.6 \\ 
  IM & 25.8 & 25.3 & 61.7 & 60.1 & 37.7 & 37.2 & 53.4 & 53.4 & 64.5 & 62.8 & 59.1 & 58.3 \\ 
     \hline  \addlinespace[.2cm]
  \multicolumn{13}{l}{\quad\textit{Scenario 2: $|A_\dev|=7$}} \\  \hline
 BH & 3.6 & 2.4 & 12.1 & 8.9 & 5.6 & 4.9 & 6.2 & 4.2 & 32.0 & 29.8 & 18.6 & 16.8 \\ 
  Holm & 1.0 & 0.9 & 11.3 & 7.9 & 3.4 & 2.6 & 3.8 & 2.4 & 11.3 & 7.9 & 7.1 & 5.1 \\ 
  IM & 7.9 & 7.8 & 61.5 & 60.1 & 16.0 & 15.8 & 40.9 & 40.9 & 61.5 & 60.1 & 47.3 & 46.4 \\ 
   \hline
\end{tabular}
}
\caption{False Discovery Rate (FDR) and family-wise Error Rate (FWER) for the three p-value combination methods from Section~\ref{subsec:combineTeststats} and the two bootstrap methods \ref{item:b2} and \ref{item:b3}. The stated values are the minimum, maximum and mean across the 224 alternative models from each scenario.} \label{tab:fdr}
\end{table}

\begin{figure}[t!]
\makebox[\textwidth][c]{
    \includegraphics[width=1.1\textwidth]{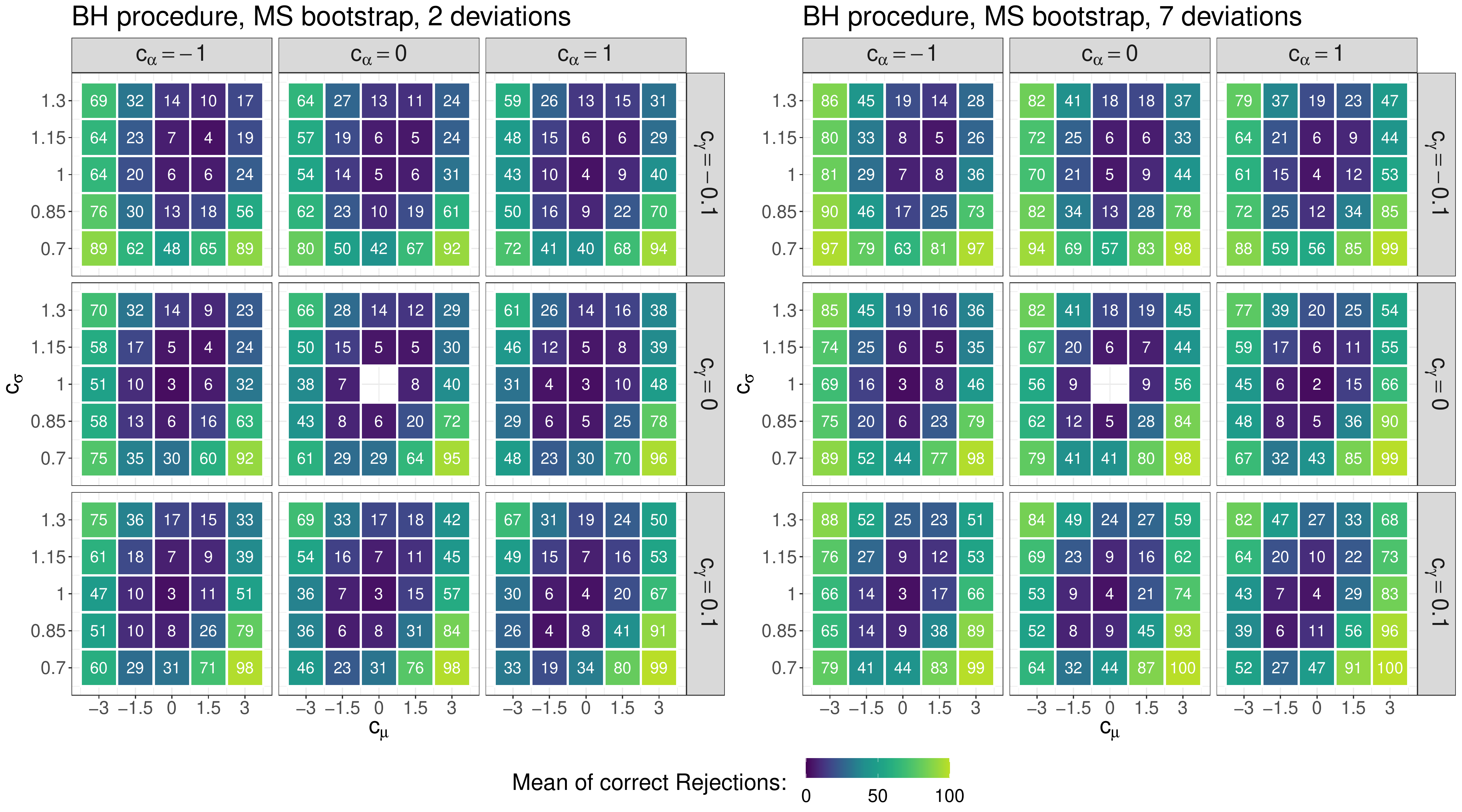} 
}
\makebox[\textwidth][c]{    
    \includegraphics[width=1.1\textwidth]{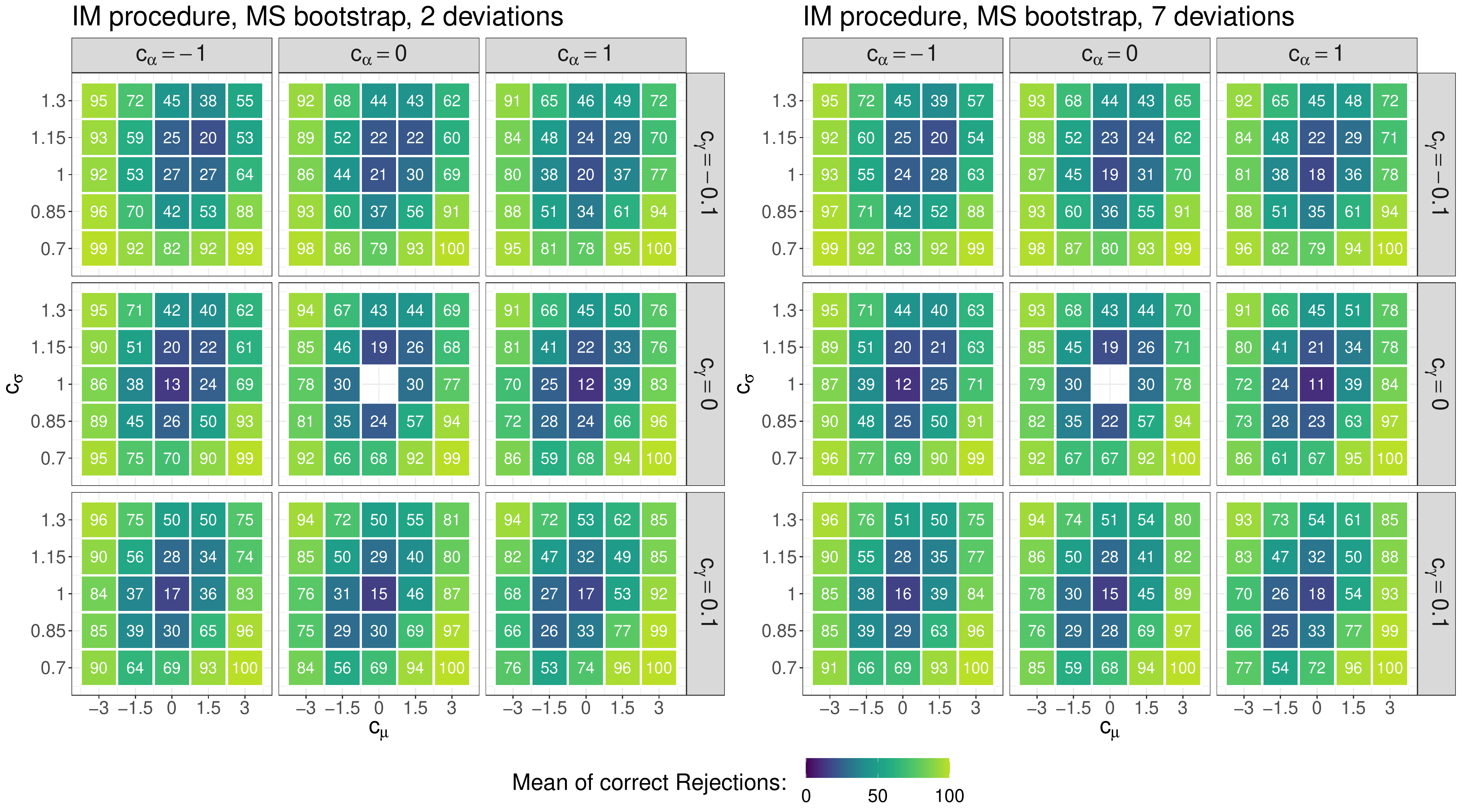}
}	
\caption{Proportion of correct rejections in $\%$ obtained with the BH procedure (upper row) and the IM procedure (lower row) at a level of 0.1, in the setting where two stations deviate from the rest (left column) or 7 locations deviate from the rest (right column), with the bootstrap procedure based on max-stable processes. 
The axis and facets are as described in Figure \ref{fig:power_ms_on16}.
}
\label{fig:bh-im-power}
\end{figure}

The power properties for procedure \ref{item:b2} combined with the BH method are shown in Figure \ref{fig:bh-im-power}. We see that the procedure is able to detect some of the deviations of the null hypothesis, with more correct rejections the stronger the deviation is. The method is particularly powerful when the location and scale parameters deviate into opposite directions, i.e. when $c_\mu > 0 $ and  $c_\sigma < 1$ or $ c_\mu < 0 $ and  $c_\sigma > 1$. There is no obvious pattern regarding the deviations of the shape and trend parameter.
Further, we analogously show the power properties of the IM method with bootstrap \ref{item:b2} in Figure \ref{fig:bh-im-power}. As expected,  this method has more power against all alternatives under consideration. However, this comes at the cost of more false discoveries, as can be seen in Table~\ref{tab:fdr}.

The results for bootstrap scheme \ref{item:b3} were very similar and are therefore not shown here, but can be found in Section \ref{app:sim} of the supplementary material. Likewise,  we omit the results for the more conservative Holm procedure, which exhibits, as expected, less power against all alternatives.
Further, we repeated the simulation study with an increased location-wise sample size of $n = 100$. As one would expect, the tests have more power in this case.

The results presented so far show that the proposed pooling methods work `as intended', since the theoretical test characteristics are well approximated in finite sample situations, and since we observe decent power properties. In practical applications however, spatial pooling of locations is usually the starting point for subsequent analyses. For instance, one may be interested in estimating return levels at the location of interest based on the data from all locations that were identified as homogeneous. Moreover, the analysis of alternative data sets like climate model data may be based on the homogeneous locations identified within the analysis of observations. 

This suggests that the methods should be examined with regard to their quality in subsequent analyses. For that purpose, we consider, as an example, the problem of return level estimation at the location of interest. The state-of-the-art method would consist of GEV fitting at the location of interest only, which results in (asymptotically) unbiased estimators that suffer from large variance. Basing the estimator on pooled regions will decrease the variance, but at the same time increase its bias if some heterogeneous locations have been wrongly identified as homogeneous. 

In particular, pooling based on a conservative testing approach like the BH procedure leads to the acceptance of many locations and thus to a large pooling area and low estimation variance. Most likely, some of the chosen locations will be violating the null hypothesis though, which yields a rather large estimation bias.
For pooling based on a more liberal rejection approach like the IM procedure, the estimation bias and variance behave exactly opposite: since the null hypotheses are more likely to be rejected, the resulting pooling sample is smaller (i.e., larger estimation variance) but `more accurate' (i.e., smaller estimation bias).

For our comparison, we consider fitting the scale-GEV model based on pooled locations that have been obtained from one of the following eight methods
\begin{align*}
m \in \{ 
\mathrm{LOI}, 
\mathrm{full},
\mathrm{MS \ IM}, 
\mathrm{MS \ Holm}, 
\mathrm{MS\ BH},
\mathrm{biv. \ IM},  
\mathrm{biv. \ Holm}, 
\mathrm{biv. \ BH} 
\}.
\end{align*}
Here, LOI refers to considering the location of interest only (no pooling), full refers to pooling all available locations, and the last six methods encode pooling based on any combination of the proposed p-value combination methods and bootstrap approaches.

For each method, we compute the maximum likelihood estimate $\hat{\bm\vartheta} =(\hat\mu, \hat\sigma, \hat\gamma, \hat\alpha)^\top$ of the scale-GEV model parameters and transform this to an estimate of the $T$-year return level (RL) in the reference climate of year $t$ by
\[ 
\widehat{\mathrm{RL}}_t(T) 
= 
G^{-1}_{(\hat \mu(t), \hat\sigma(t), \hat\gamma)}(1-1/T),
\]
where $\hat \mu(t) = \hat \mu \exp(\hat \alpha \gmst(t) / \hat \mu)$ and $\hat \sigma(t) = \hat \sigma \exp(\hat \alpha \gmst(t) / \hat \mu)$ and where $G$ is the cumulative distribution function of the GEV-distribution, see Equation \eqref{eq:gevcdf}.
Now, in our simulation study, we know that the true value of the target RL is given by $\mathrm{RL}_t(T) =  G^{-1}_{(\mu_0(t), \sigma_0(t), \gamma_0)}(1 - 1/T)$ with 
\[ 
\mu_0(t) = 20\exp\left( \frac{1.5 \gmst(t)}{20}\right), \, 
\sigma_0(t) = 5.5\exp\left( \frac{1.5 \gmst(t)}{20}\right), \, 
\gamma_0 = 0.1.
\] 
From the 500 replications we can therefore compute the empirical Mean Squared Error (MSE) of method $m$ as 
\[ 
\mathrm{MSE}(m) =  \frac{1}{500} \sum_{j = 1}^{500} \left(\widehat{\RL}^{(m, j)}_t(T) - \RL_t(T) \right)^2 , 
\]
where $ \widehat{\RL}^{(m, j)}_t(T)$ denotes the estimated RL obtained in the $j$-th replication with method $m$. Note that we have suppressed the MSE's dependence on $T$ and $t$ from the notation.

\begin{figure}[t!]
\makebox[\textwidth][c]{
    \includegraphics[width=1.1\textwidth]{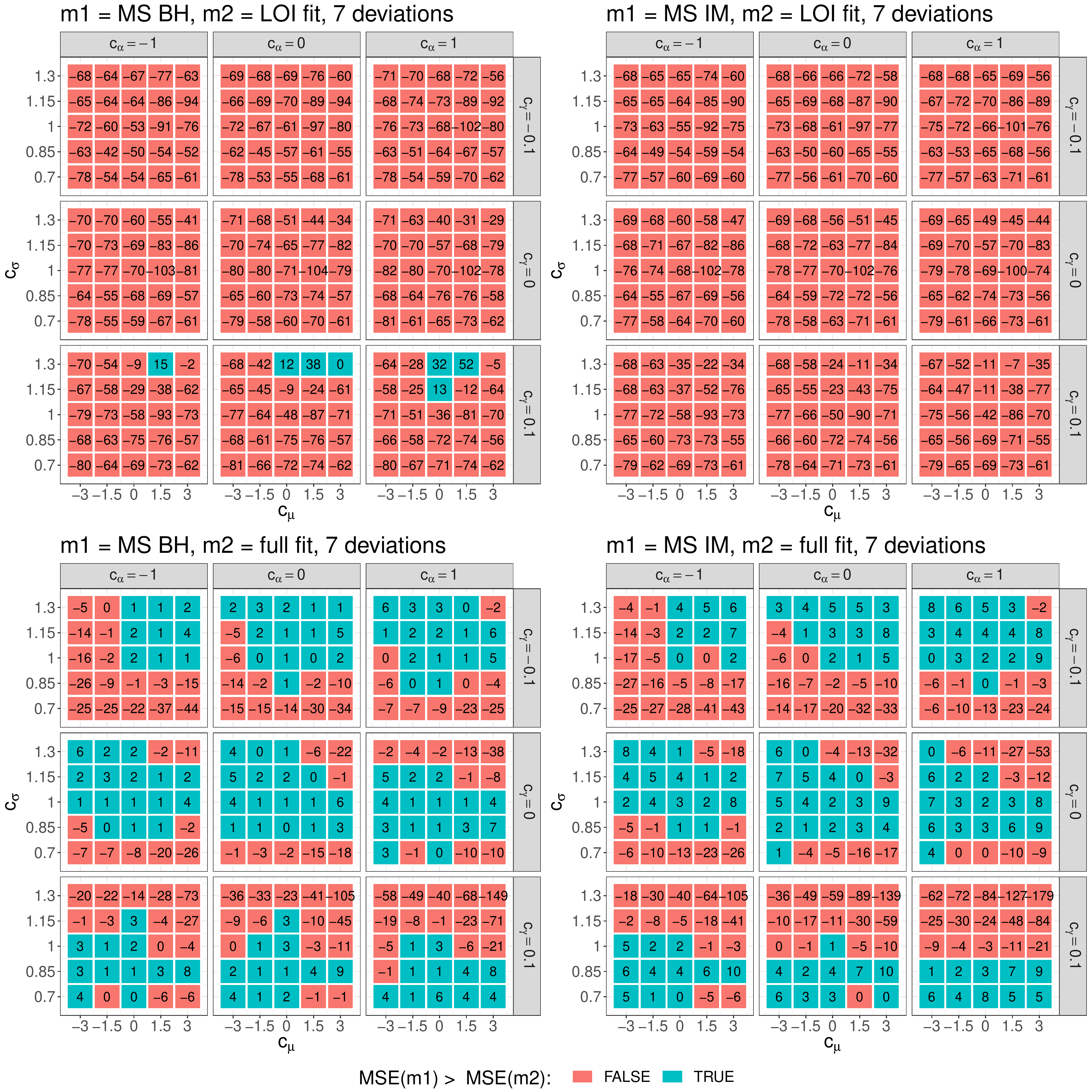}
}	
\caption{Comparison of MSEs of $\RL_{2021}(100)$ obtained for different choices of the method $m$, in the setting where $|A_\dev| = 7$. Shown are the differences $\mathrm{MSE}(m1) - \mathrm{MSE}(m2)$ with $m1$ and $m2$ as indicated in the plot title. Negative values (red) therefore indicate a lower MSE for the method mentioned first, and vice versa for positive values. 
The axis and facets are as described in Figure \ref{fig:power_ms_on16}.
} 
\label{fig:MSErl_MS}
\end{figure}

In Figure \ref{fig:MSErl_MS} we compare MSEs of the 100-year RL with reference climate as in year 2021, which is given by $\RL_{2021}(100) = 55.87$, by plotting the difference $\mathrm{MSE}(m_1) - \mathrm{MSE}(m_2)$  with $m_1 \in \{\mathrm{MS\, BH, MS\, IM}\}$   and $m_2 \in \{\mathrm{full}, \mathrm{ROI}\}$ as obtained for the setting where $|A_\dev| = 7$. 
The plots reveal that both the MS BH and the MS IM method are superior to the the LOI fit for almost all scenarios. Comparing the two methods to the full fit reveals that there are certain scenarios for which the full fit performs substantially worse, mostly when the shape and scale parameter deviate towards the same direction for the alternatives.
For those scenarios where the full fit outperforms the two methods, the discrepancy is not very large, with the BH method performing slightly better than the IM method.

\begin{figure}[t!]
\makebox[\textwidth][c]{
    \includegraphics[width=1.1\textwidth]{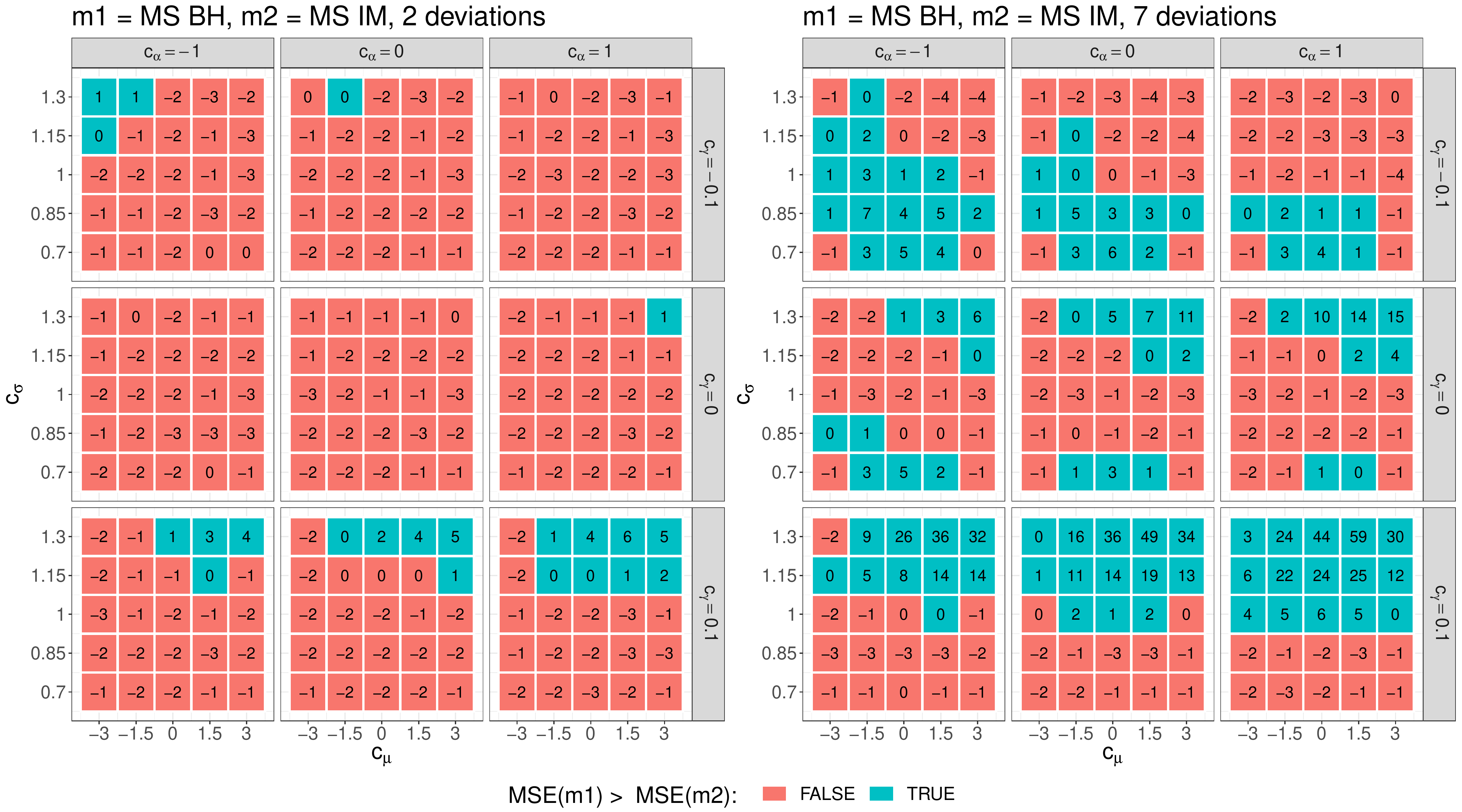}
}	
\caption{Comparison of MSEs of $\RL_{2021}(100)$ in the setting where $|A_\dev|  = 2$ (left) and $|A_\dev|  = 7$ (right). Shown are the differences  MSE(MS BH) $-$ MSE(MS IM). Negative values (red) therefore indicate a lower MSE for the BH method, while positive values (blue) indicate a lower MSE for the IM method. 
The axis and facets are as described in Figure \ref{fig:power_ms_on16}.
\label{fig:MSErl_MS_bhim} }
 \end{figure}

A comparison between MS BH and MS IM is shown in Figure \ref{fig:MSErl_MS_bhim} for $|A_\dev| \in \{2,7\} $. 
The results reveal that the BH method slightly outperforms the IM method in the case $|A_\dev| = 2$ for almost all alternative scenarios. In case $|A_{\dev}|=7$, the results are quite mixed, with the IM method becoming clearly superior when the shape, scale and location parameters deviate jointly to the top. In all other scenarios, the differences are only moderate, sometimes favoring one method and sometimes the other. Corresponding results for the bootstrap methods based on bivariate extreme value distributions are very similar and therefore not shown. Further, the results were found to be robust against the choices of $t = 2021$ and $T = 100$ that were made here for the return level estimation. 

Overall, the results suggest the following practical recommendation: if the full sample is expected to be quite homogeneous a priori (for instance, because it was built based on expert knowledge), then estimation based on BH-based pooling is preferable over the other options (LOI, the full and the IM-based fit). If one expects to have a larger number of heterogeneous locations, it is advisable to apply the IM procedure (or any other liberal procedure), which likely rejects most of the heterogeneous locations and hence reduces the bias. In general, the liberal behavior of IM-based pooling suggests its use when it is of highest practical interest to obtain a pooled region that is as homogeneous as possible (as a trade-off, one has to accept that the region is probably much smaller than the initial full region).

\section{Severe flooding in Western Europe during July 2021 revisited} 
\label{sec:case}

We illustrate the new pooling methods in a case study by revisiting the extreme event attribution study for the heavy precipitation event that led to severe flooding in Western Europe during July 2021, see \cite{Kre21,Tra22}. In that study, observational data were pooled together based on expert knowledge and on ad hoc tests, with the ultimate goal of assessing the influence of human-made climate change on the likelihood and severity of similar events in Western and Central Europe.

\begin{figure}[t!]
    \centering
    \begin{minipage}{.5\textwidth}
        \centering
        \includegraphics[width=.99\textwidth]{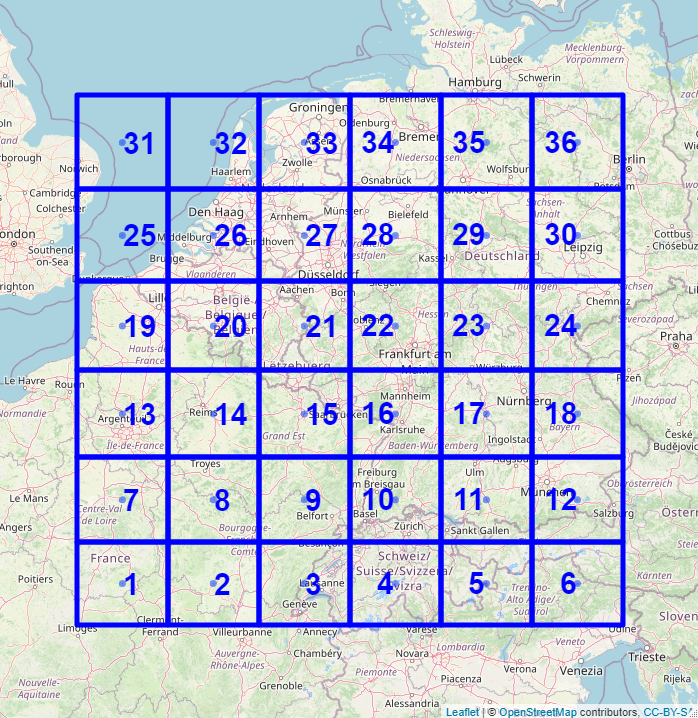}
    \end{minipage}%
    \begin{minipage}{0.5\textwidth}
        \centering
        \includegraphics[width=.99\textwidth]{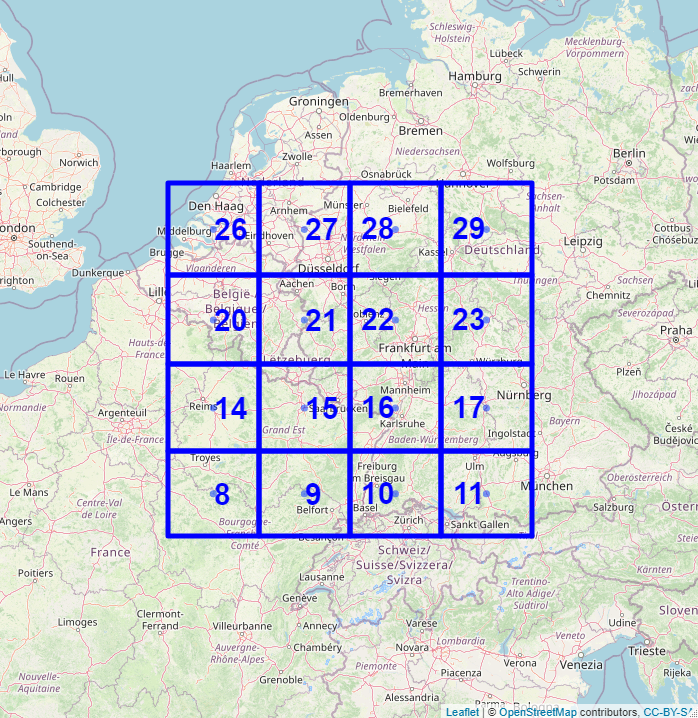}
    \end{minipage}
        
    \caption{\label{fig:grid_plain} Regions analysed within this case study and the respective numbering used here. The data consists of April-September block maxima of tile-wise averaged daily precipitation sums (RX1day) from 1950-2021.}
\end{figure}

The full region under investigation in \cite{Kre21,Tra22} consists of sixteen $(2.0^\circ \times 1.25^\circ)$ (i.e. about $(140\,\mathrm{km} \times 140 \,\mathrm{km})$)
grid cells reaching from the northern Alps to the Netherlands, see Figure 5 in \cite{Kre21} or the right-hand side of Figure \ref{fig:grid_plain}. Two of the 16 locations were rejected in that study due to expert knowledge and too large deviations in fitted GEV-parameters (regions 17 and 11 of Figure \label{ref:grid_plain}). Among other things, our illustrative application of the methods explained above will reveal that grid cell 11 has been rightfully dismissed, while grid cell 17 might have been considered homogeneous. Further, there is no clear evidence that any other grid cell that has been declared homogeneous should rather be considered non-homogeneous.

For illustrative purposes, we apply our methods to two different initial areas:
\begin{compactenum}
    \item[(A)] An area consisting of $6 \times 6$ grid cells covering a large part of Western/ Central Europe, as shown in Figure \ref{fig:grid_plain} on the left.
     \item[(B)] The original $4 \times 4$ grid cells from \cite{Kre21} as shown in Figure \ref{fig:grid_plain} on the right.
\end{compactenum}
Note that homogeneity for the 20 grid cells at the boundary of the larger area in (A) has been dismissed based on expert knowledge in \cite{Kre21}; the larger area is included here for illustrative purposes only.

The data used throughout the study consists of April-September block-maxima of tile-wise averaged 1-day accumulated precipitation amounts of the E-OBS data set (\cite{cornes2018ensemble}, Version 23.1e).
In both cases, the grid cell with label 21 is the one of primary interest, since it is the one containing the target location of the study, i.e., the region that accumulated the highest precipitation sum and experienced the largest impacts during the flooding of July 2021. 
The time series are shown in Figure~\ref{fig:obs_with_trend} in the supplementary material. There, we also plot values of 
$\hat\mu(t) = \hat\mu\exp\left( \hat\alpha\gmst(t)/\hat\mu  \right)$ obtained from different data sets: once from 
data of
location 21 only, once from data of the respective location only, and once from the pooled data of the respective pair $(21, d)$ for $ d \in \{ 1, \ldots, 36\} \setminus\{21\}$. 

We apply the two proposed bootstrap procedures to areas (A) and (B). Note that the raw p-values obtained with the bootstrap based on bivariate extreme value distributions should be very similar (or even identical when using the same seed for random number generation) for those grid cells that appear in both areas, while they may differ to a greater extent for the MS bootstrap. 
This is because the p-value for a given pair obtained with the bivariate bootstrap procedure only depends on the observations of the pair, while the respective p-value obtained with the MS bootstrap also depends on the spatial model that was fitted to the whole area.
However, even if the raw p-value of a given pair obtained for setting (B) coincides with the raw p-value obtained for setting (A), the adjustment for multiple testing can lead to slightly different rejection decisions of the pair at a given level $\alpha$.
The bootstrap procedures are applied with $B = 2000$ bootstrap replications.

We start by discussing the results of the application to the larger grid in (A). 
Recall that, for a given significance level $\alpha$, one rejects the null hypothesis for all grid cells whose p-value is smaller than $\alpha$. To visualise the results, we therefore shade the grid cells according to the magnitude of their (adjusted) p-value. Here, we divide the colour scale into three groups: 
$[0, 0.05], (0.05, 0.1]$ and $ (0.1, 1]$, with a dark red tone assigned to the first group, a brighter red tone for Group 2 and an almost transparent shade for Group 3.  
This allows us to see the test decisions for significance levels of $\alpha \in \{ 0.05, 0.1\}$: 
when the significance level is chosen as $\alpha = 0.1$, all tiles with a reddish shade are rejected, while when working with a level of $\alpha = 0.05$ only tiles shaded in the dark shade are rejected.

\begin{figure}[t!]
    \centering
    \begin{minipage}{.5\textwidth}
        \centering
        \includegraphics[width=.99\textwidth]{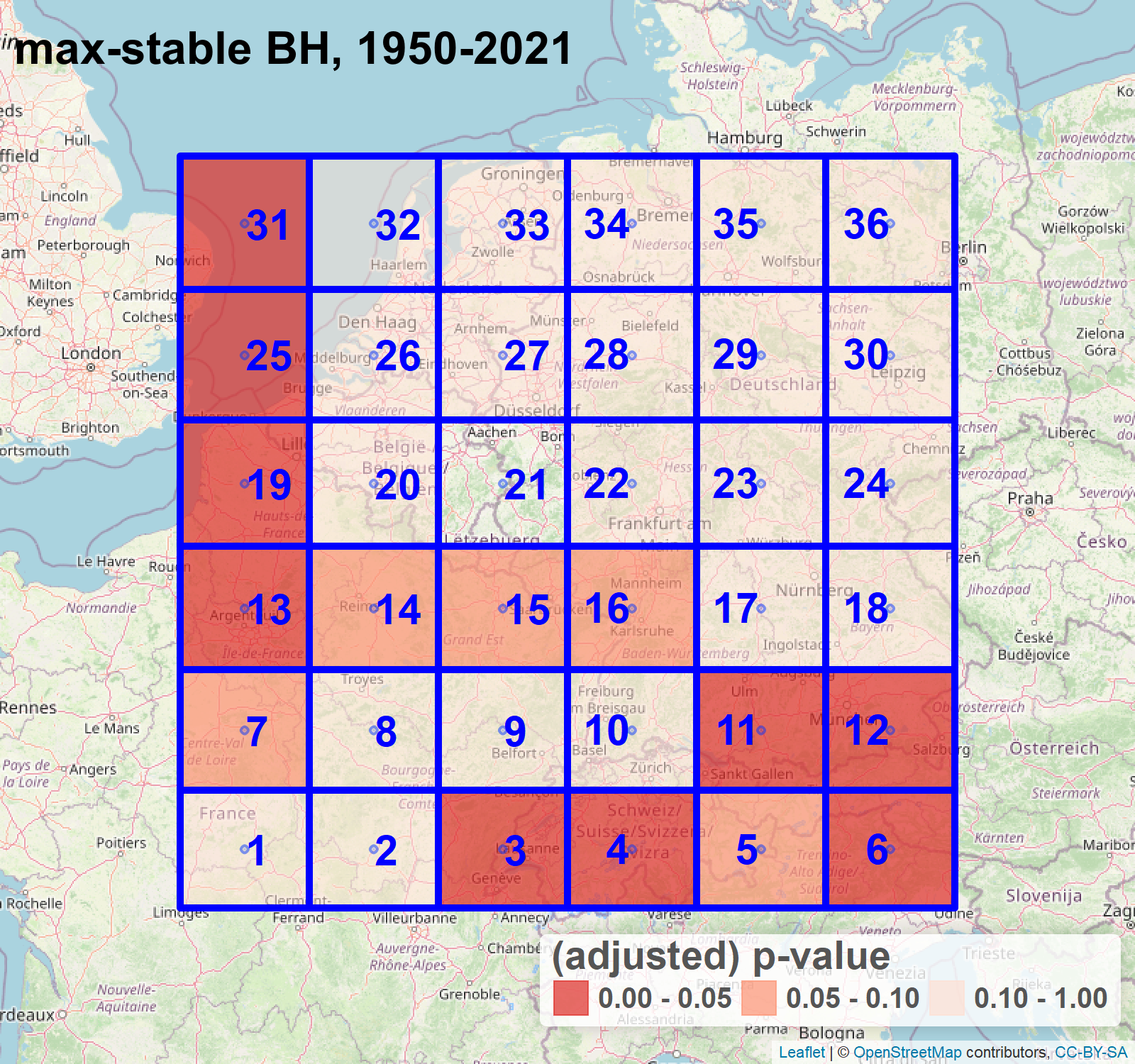}
    \end{minipage}%
    \begin{minipage}{0.5\textwidth}
        \centering
        \includegraphics[width=.99\textwidth]{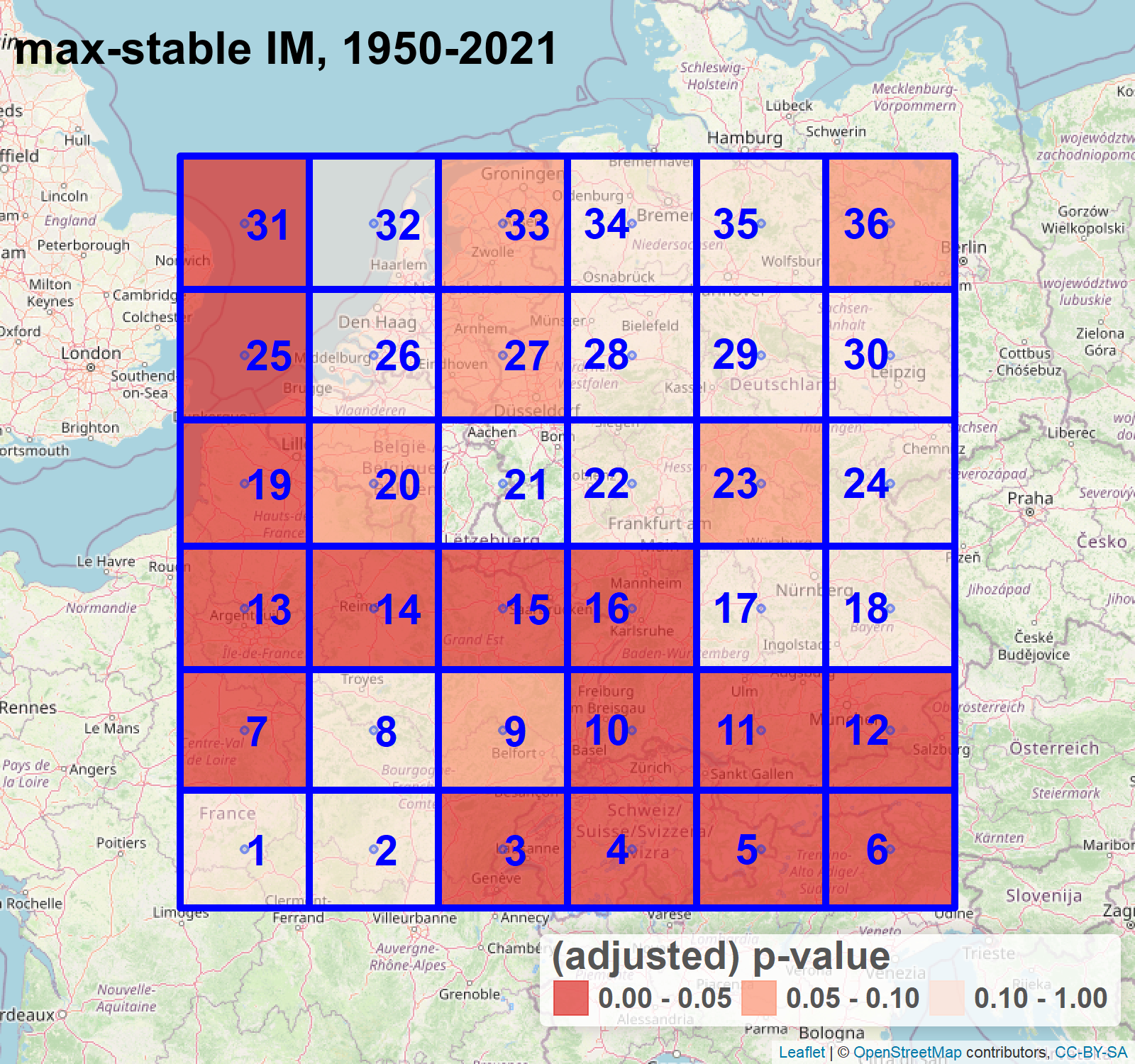}
    \end{minipage}
      \begin{minipage}{.5\textwidth}
        \centering
        \includegraphics[width=.99\textwidth]{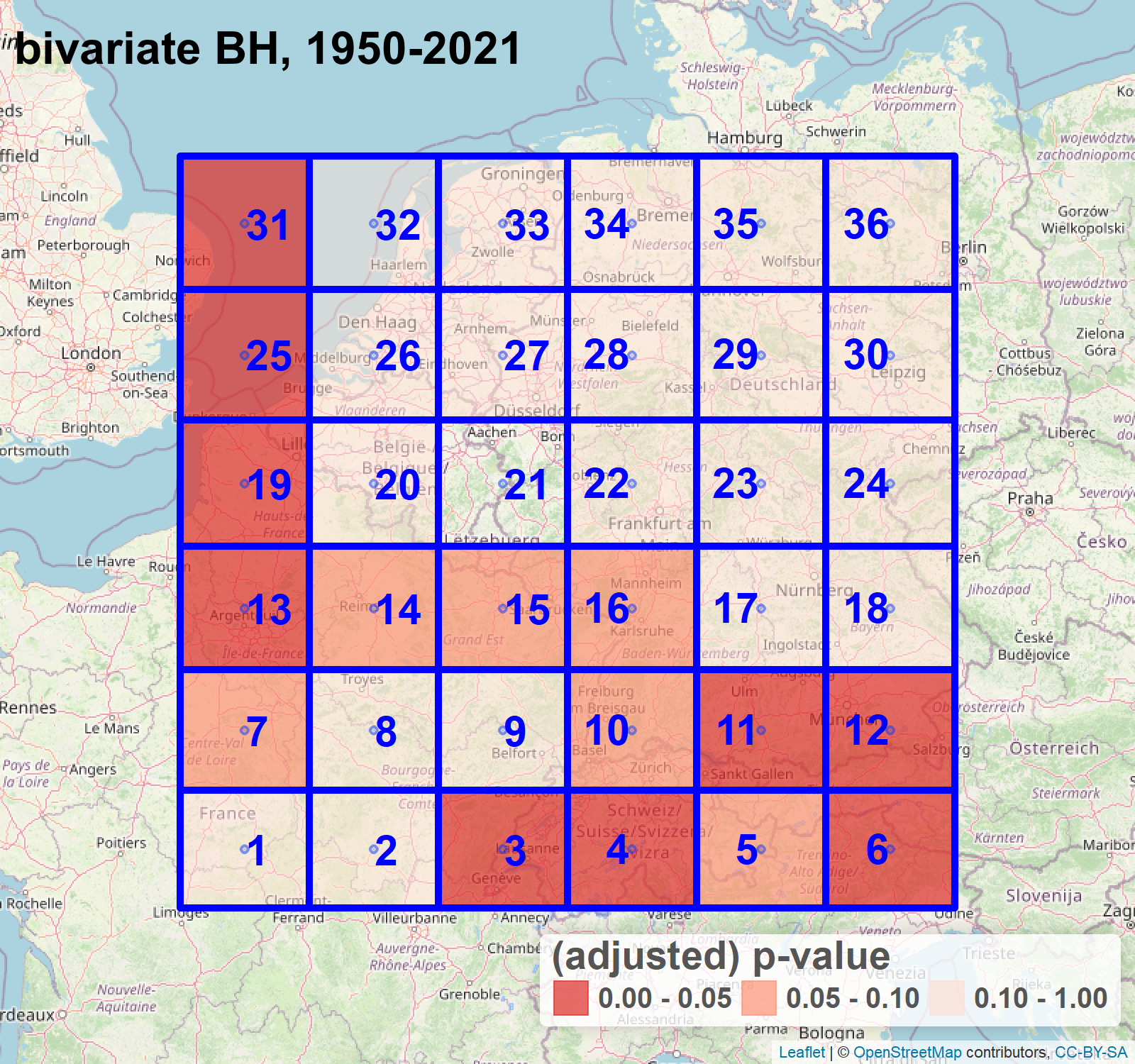}
    \end{minipage}%
    \begin{minipage}{0.5\textwidth}
        \centering
        \includegraphics[width=.99\textwidth]{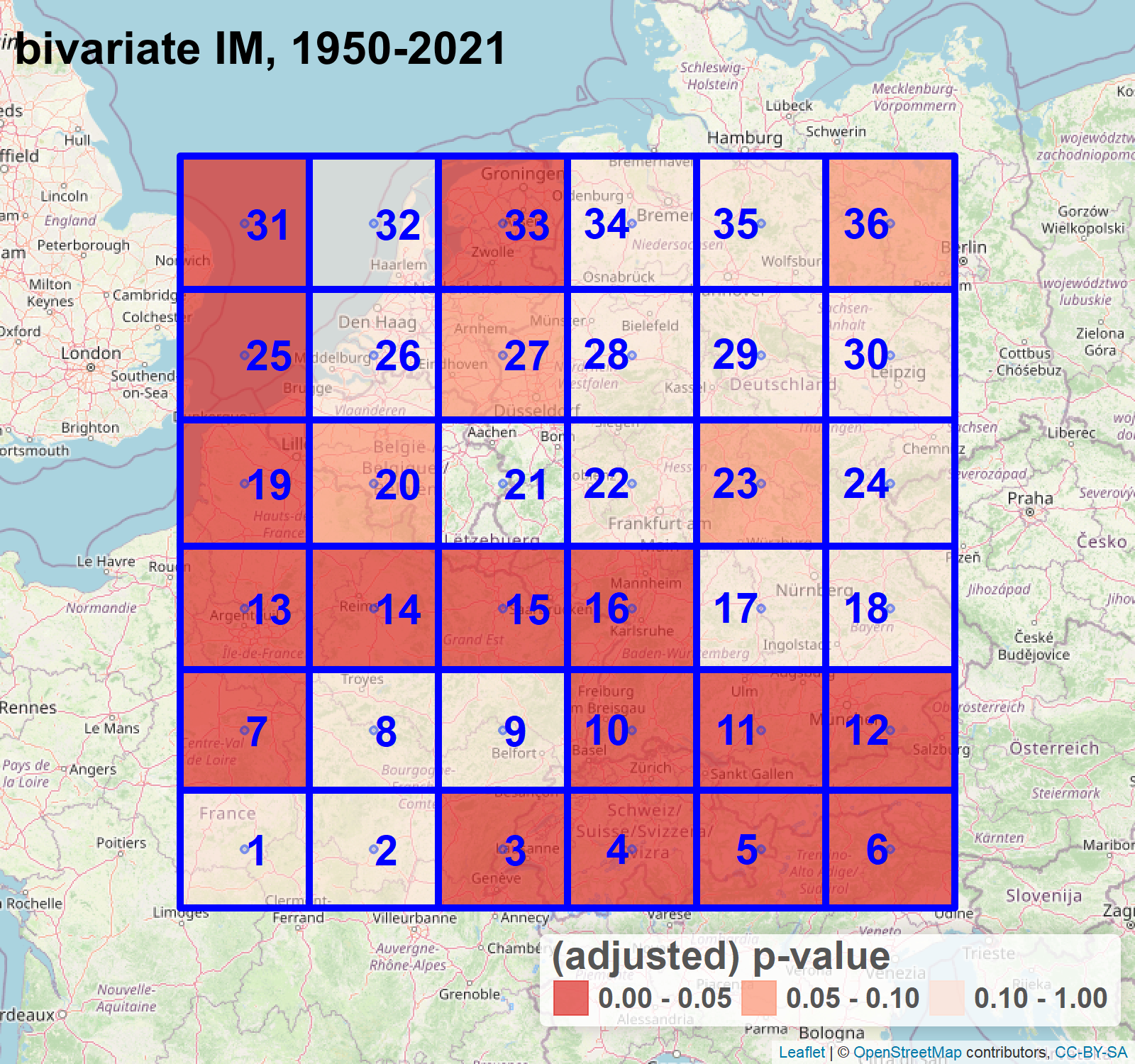}
    \end{minipage}
    \caption{\label{fig:pval36} (Adjusted) $p$-values obtained with the BH (left) and the IM (right) method on the $6\times 6$ grid, with the bootstrap based on max-stable processes (top row) and the bootstrap based on bivariate extreme value distributions (bottom row). }
\end{figure}

Results for both the conservative BH procedure and the liberal IM procedure are shown in Figure \ref{fig:pval36}. For completeness, results on Holm's method, which is even more conservative than BH, as well as the BH and IM p-values themselves can be found in the supplementary material, Tables~\ref{tab:p-case-biv} and \ref{tab:p-case-ms}. 
One can see that, for a given rejection method (i.e. BH or IM), the MS and bivariate procedures mostly agree on the rejection decisions that would be made at a level of 10\% (compare the rows of Figure \ref{fig:pval36} to see this).
The same holds when working with a significance level of 5\%. 

Further, as expected, the IM method rejects more hypotheses than the BH method. However, according to the results of the simulation study, it is quite likely that at least one of these rejections is a false discovery.

\begin{figure}[t!]
    \centering
    \begin{minipage}{.5\textwidth}
        \centering
        \includegraphics[width=.99\textwidth]{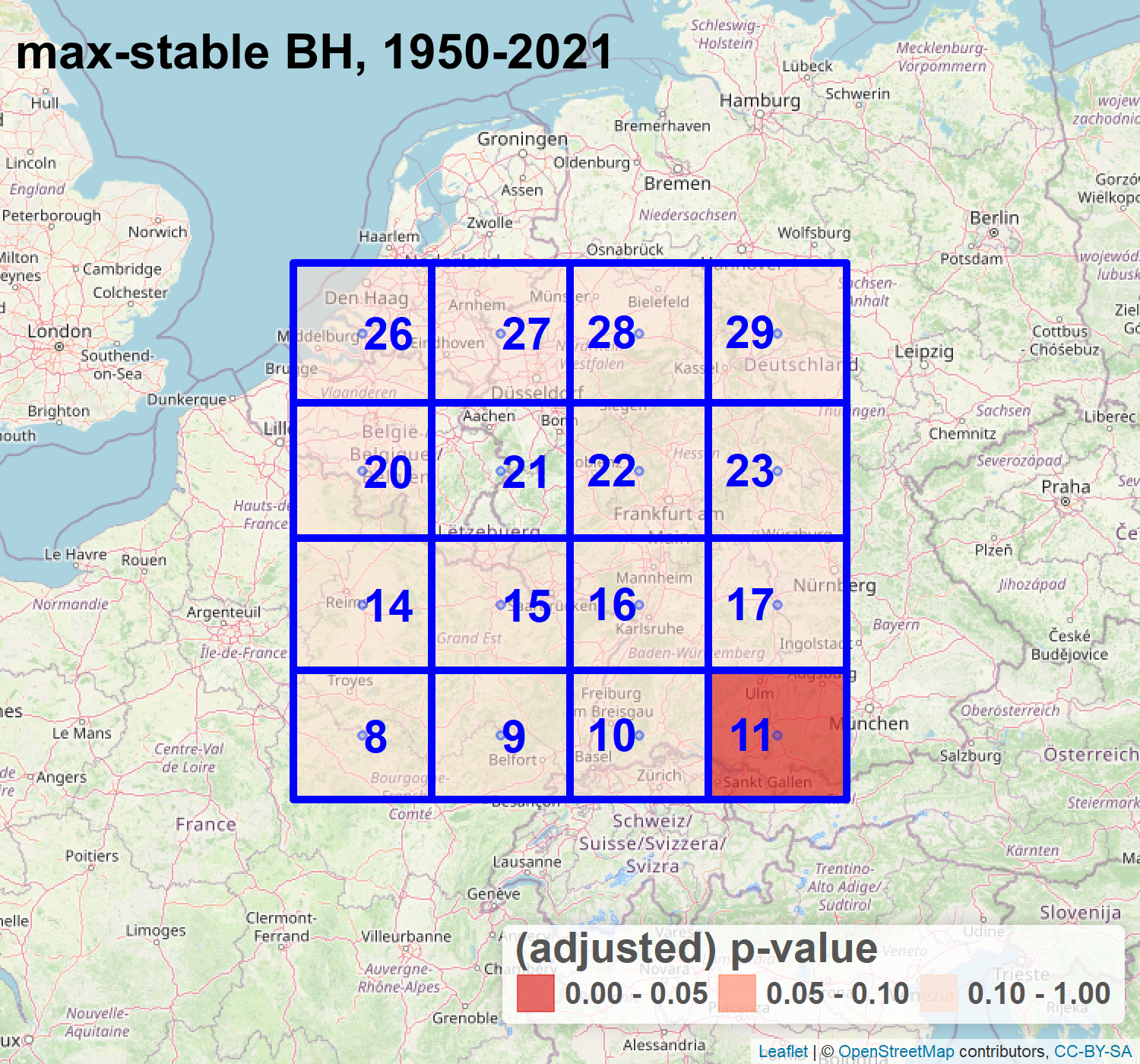}
    \end{minipage}%
    \begin{minipage}{0.5\textwidth}
        \centering
        \includegraphics[width=.99\textwidth]{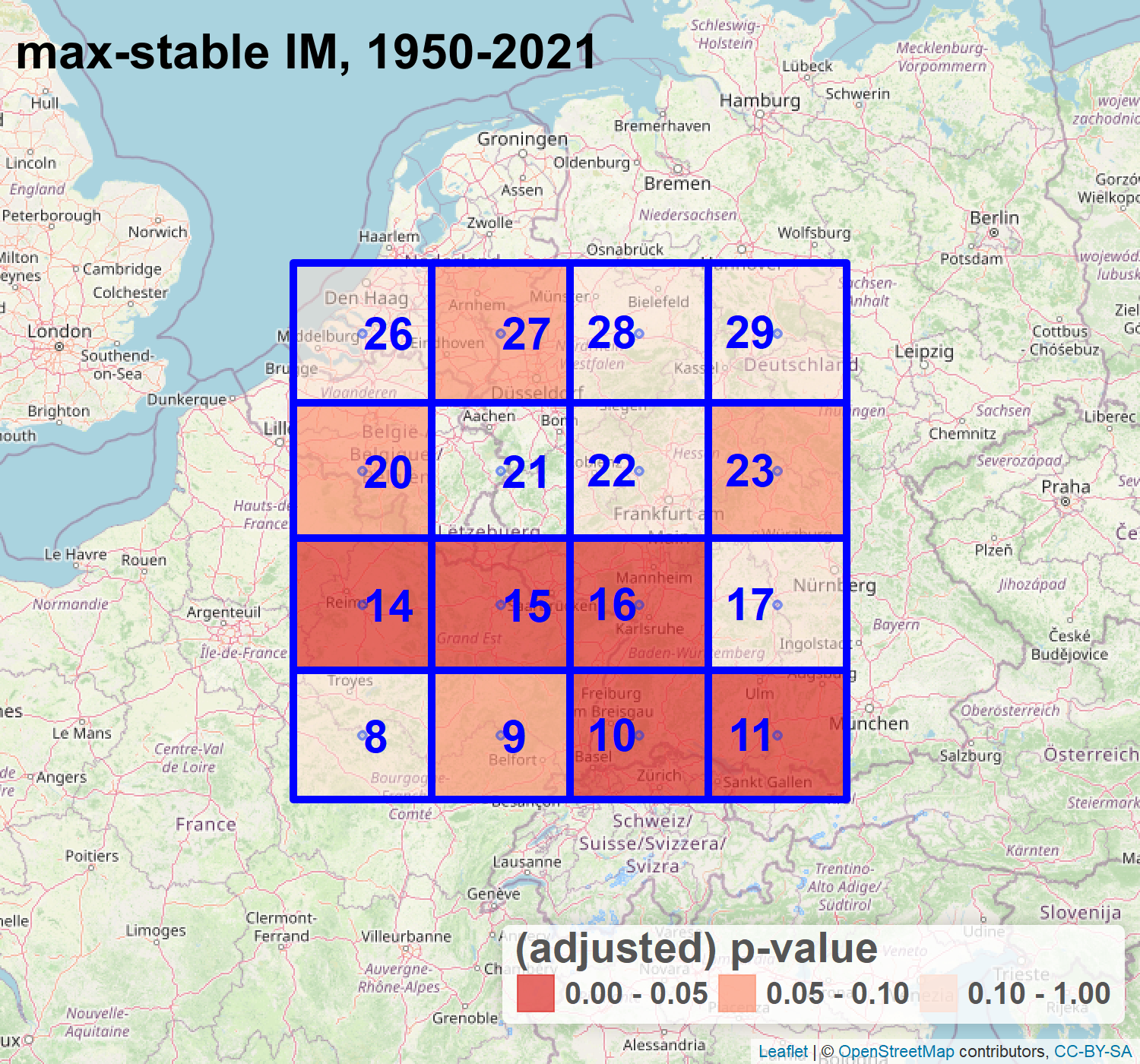}
    \end{minipage}
    \begin{minipage}{.5\textwidth}
        \centering
        \includegraphics[width=.99\textwidth]{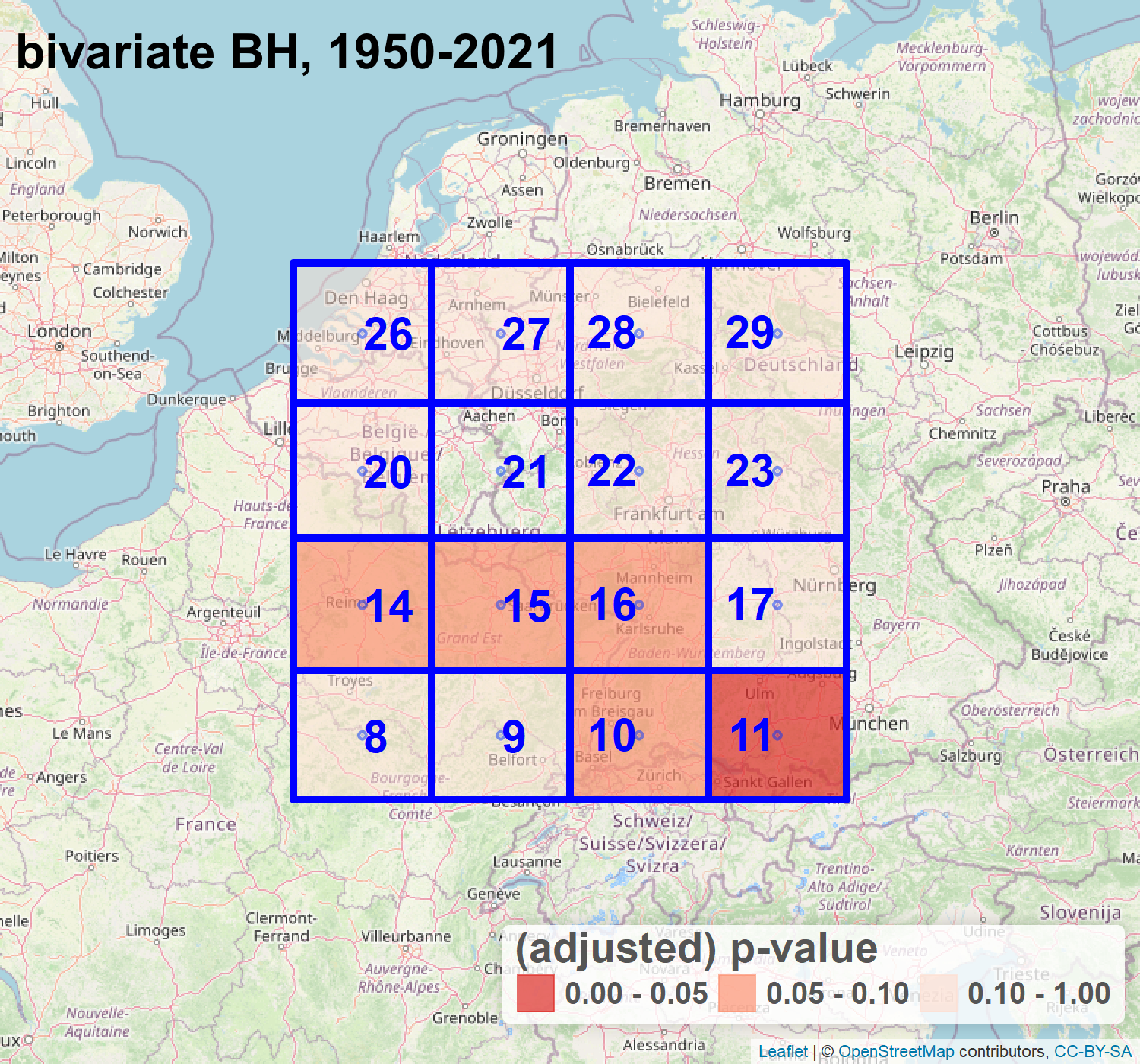}
    \end{minipage}%
    \begin{minipage}{0.5\textwidth}
        \centering
        \includegraphics[width=.99\textwidth]{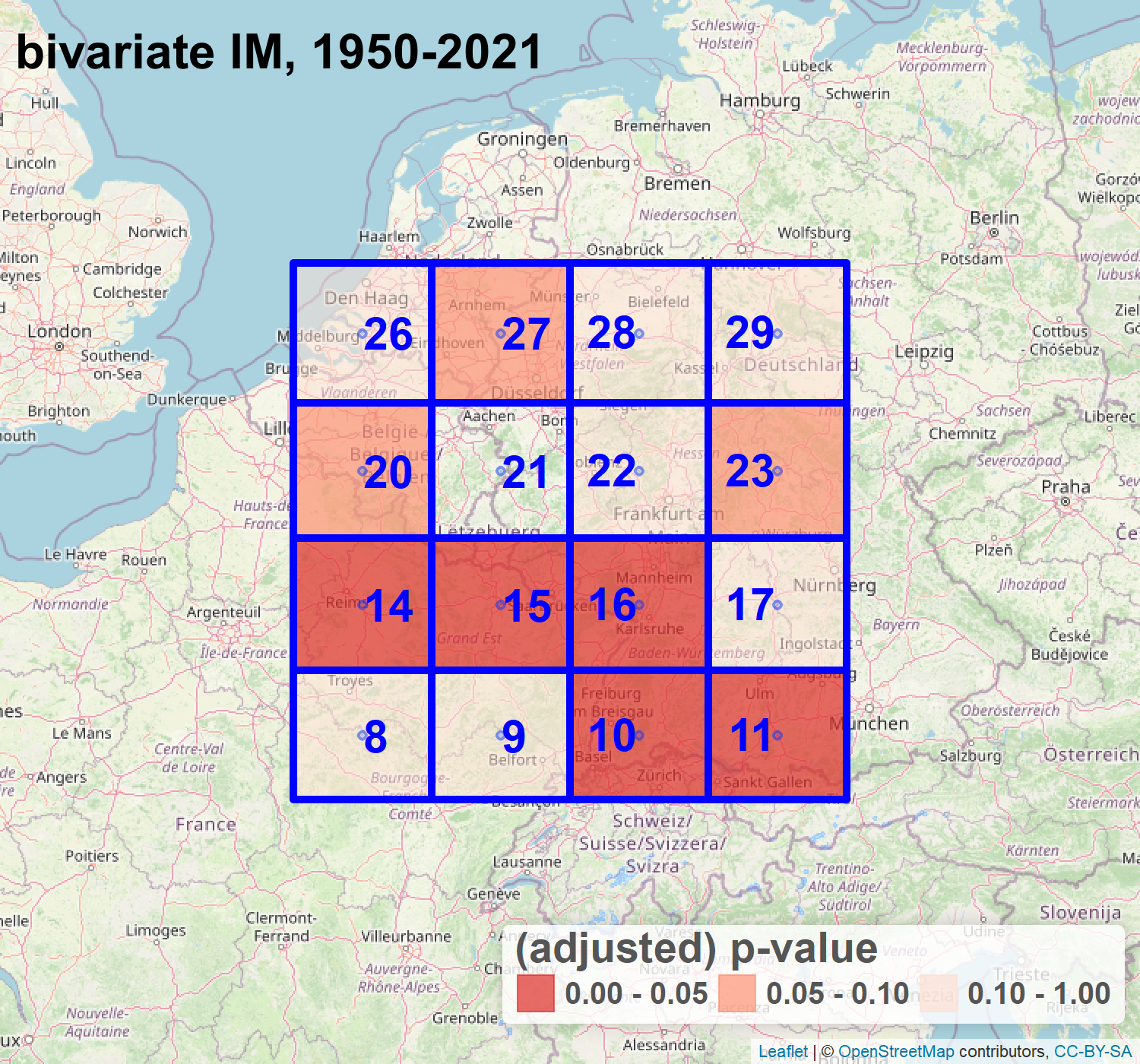}
    \end{minipage}
    \caption{\label{fig:pval16} Adjusted $p$-values obtained with the BH (left) and the IM (right) method on the $4\times 4$ grid, obtained with the bootstrap based on max-stable processes (top row) and the bootstrap based on bivariate extreme value distributions (bottom row).}
\end{figure}

Analogous results for the $4\times 4$ grid in (B) are shown in Figure \ref{fig:pval16}. 
As discussed above, except for the MS BH method, the results are consistent with the results obtained for the  $6\times 6$ grid in the sense that 
for those locations which are contained in both grids, the locations with
p-values of critical magnitude ($< 10\%$) coincide (compare the plot in the upper right corner of Figure~\ref{fig:pval16} to the plot in the upper right corner of Figure~\ref{fig:pval36} to see this for the MS IM method, and similar for the other methods).
For the MS BH method, grid cells 10, 14, 15, and 16 are not significant anymore at a level of 10 \%, but we recorded an adjusted p-value of 0.106 for those four grid cells, so this is a rather tight decision.
 The p-values obtained for the $4\times 4$ grid can be found in Table~\ref{tab:p-case-16} in the supplementary material.

Let us now move on to the interpretation: considering the larger grid first, some grid cells for which the characteristics of extreme precipitation are different (according to expert opinion) from the grid cell of the target location are detected as heterogeneous. These rejected grid cells are located along the coast and in the mountainous terrain.
Comparing the results with \cite{Kre21, Tra22}, we observe that grid cell~11 has been rejected in their study based on expert knowledge. For grid cell~17, however, we do not detect any statistical evidence that the probabilistic behavior of extreme precipitation is different from the grid cell of the target location, even when applying the liberal IM procedure. We would like to stress though that non-rejection of a null hypothesis does not provide any evidence of the null hypothesis, even when ignoring the multiple testing issue. Hence, expert knowledge that leads to rejection should, in general, outweigh any statistical non-rejection. This particularly applies to the eastern (continental) grid cells in the larger 6$\times$6-grid, which can be influenced by heavy precipitation caused by different synoptic situations compared to the target region.

Moreover, as the results for locations 10, 14, 15, and 16 showed some discrepancy across the different testing procedures, we suggest that the final decision on the exclusion or inclusion of these locations in a spatial pooling approach should be based on expert knowledge of the meteorological characteristics,
and the willingness to trade possible bias for variance (with a possibly larger bias when including the locations -- note that statistical evidence against homogeneity in the bivariate extreme value distribution-based bootstrap is only weak, and wrongly declaring  the regions as homogeneous is possibly not too harmful).
The same holds for locations 9, 20, 23 and 27 for which only the IM method yielded p-values between 5\% and 10\%. Again, these rather small p-values could be subject to false discoveries though, and since the heterogeneity signal is also not too strong, there is no clear evidence that these need to be excluded from pooling.

For a last evaluation of results from pairwise tests, we estimated the 100-year RLs in the reference climate of the year 2021, i.e. with reference value $\gmst(2021) = 0.925 ^\circ C$, on five different data sets obtained from the $4\times 4$ grid. Here, we use the data sets consisting of data from
\begin{compactitem}
\item the location of interest only
\item pooling those grid cells suggested by the results of the case study (i.e., all cells but 11, or all cells but 10, 11, 14, 15, 16) or expert opinion (i.e., all cells but 11, 17)
\item  pooling all grid cells of the $4\times 4$ grid.
\end{compactitem}
The results can be found in Table~\ref{tab:cs_rl} and reveal that excluding cell 11 has a clear effect on the estimated RL. 
Ex- or including grid cell 17 once 11 is excluded does not have a large effect, while excluding cells 10, 14, 15 and 16 additionally to cell 11 has a moderate effect. 

\begin{table}[ht]
\centering
\begin{tabular}{rrrrrr}
 \hline
 \rule{0pt}{3ex}
 cells excluded & $\hat\mu$ & $\hat\sigma$ & $\hat\gamma$ & $\hat\alpha$ & $\widehat\RL_{2021}(100)$ \\ 
  \hline
 none &  20.37 &  5.80  & 0.1039 & 1.50 & 58.43   \\ 
11 & 20.01 & 5.44 & 0.0676 & 1.45 & 52.74 \\ 
 11, 17 &  20.01 & 5.40 & 0.0760 & 1.29 & 52.82\\ 
  10, 11, 14, 15, 16 & 19.90 & 5.41 & 0.0484 & 1.79 & 51.93  \\ 
 all but 21 &  21.92 & 6.08 &  0.0634 & $-0.00$ &  54.37 \\
   \hline
\end{tabular}\caption{\label{tab:cs_rl} 
Estimated parameters and estimate of $\RL_{2021}(100)$ obtained when pooling all grid cells but the ones given in the first column.
}
\end{table}

Finally, we would like to mention that similar results were obtained when applying the BH test procedures to all triples containing the pair of grid cells $(20, 21)$, i.e., the extended target region considered in the study of~\cite{Kre21,Tra22}, consisting of those regions in Germany and Belgium affected worst by the July 2021 flooding.

\section{Extensions} \label{sec:extensions}

In this section, we discuss how to estimate region-wise return levels under homogeneity assumptions (Section~\ref{sec:regrl}). We also propose two possible extensions of the pooling approach from the previous sections to other hypotheses (Section~\ref{subsec:if}) or other underlying model assumptions (Section~\ref{subsec:smooth}).

\subsection{Estimation of regional return levels and return periods}
\label{sec:regrl}

As pointed out in \cite{Kre21,Tra22} among others, an estimated return period (RP) of $T$ years for a given event and in a fixed reference climate (e.g., the preindustrial climate), obtained based on a fit of the GEV distribution to a pooled homogeneous sample, has the following interpretation: for each fixed location/tile within the region, one can expect one event of the same or larger magnitude within $T$ (imaginary) years of observing the reference climate. We refer to this quantity as the local return period.
Obviously, one would expect more than one event of similar magnitude happening at \textit{at least one of the locations} of the pooling region. 
Likewise, for a given $T$, one would expect a higher $T$-year return level for the whole region. The latter corresponds to the value that is expected to be exceeded only once in $T$ years at \textit{at least one of the locations}.

Mathematically, using the notation from Section~\ref{subsec:model}, the exceedance probability of value $r$ at at least one among $D\ge2$ locations in the reference climate corresponding to year $t$ is given by 
\begin{align*}
p_t(r) 
= 
P\Big(\exists\, j \in\{1, \dots, D\}: M_j^{(t)} \ge r \Big)
&= 
P\Big(\max_{ d = 1, \ldots, D} M_d^{(t)} \geq r \Big),
\end{align*}
such that the return period for event $r$ of the region is $\mathrm{RP}_{t}^{\mathrm{reg}}(r) = \frac{1}{p_t(r)}$. Further, the $T$-year return level of the region in the climate corresponding to year $t$ is the minimal value $\RL_{t}^{\mathrm{reg}}(T)$ for which
\[ 
P\Big(\max_{ d = 1, \ldots, D} M_d^{(t)} \geq \RL_{t}^{\mathrm{reg}}(T) \Big) \leq \frac{1}{T} \]
holds.
Both quantities could be computed (exactly) if one had access to the distribution of $\max_{d=1, \ldots, D} M_{d}^{(t)}$. For example, if the random variables $M_{d}^{(t)}, \ d = 1, \ldots, D$ were independent, $p_t(r)$ could be further simplified to 
\begin{align*}
    p_t(r) 
    = 1- P\Big(\max_{ d = 1, \ldots, D} M_d^{(t)} \leq r \Big) 
    = 1 - (G_{(\mu(t), \sigma(t), \gamma)}(r))^D,
\end{align*}
where $G$ is the distribution function of the GEV distribution and where $\mu(t), \sigma(t)$ and $\gamma$ denote the parameters at reference climate of year $t$ from Equation~\eqref{eq:ms0} under the homogeneity assumption from Equation \eqref{eq:h0}.

The locations are, however, usually not independent in applications. In the following, we propose a simulation-based estimation method that involves max-stable process models to account for the spatial dependence. As before, the \texttt{R} package \texttt{SpatialExtremes} \citep{SpatialExtremes} allows for fitting and simulating max-stable process models.

\begin{algorithm}{(Simulation-based estimation of the regionwise RL and RP)}
    \begin{compactenum}[(1)]
        \item Fit the scale-GEV parameters to the pooled homogeneous sample, resulting in the parameter vector $\hat{\bm\vartheta} = (\hat\mu, \hat\sigma, \hat\gamma, \hat\alpha)^\top$.
        \item Transform the margins of the pooled data to approximately unit Fréchet by applying transformation from Equation \eqref{eq:gev2frech} with the parameter estimate from Step 1. Then fit several max-stable process models to the obtained data and choose the best fit according to the information criterion CLIC.
        \item Replicate for $b = 1, \ldots, B$ the following steps:
        \begin{compactenum}[(i)]
            \item Generate one random observation $(y_{1,b}^{(t), \ast}, \ldots,y_{D,b}^{(t), \ast})$ from the chosen max-stable process model.
            \item Transform the margins to GEV margins, by applying the transformation in \eqref{eq:frech2gev} with parameters as estimated in Step 1, resulting in the observation $(m_{1,b}^{(t), \ast}, \ldots,m_{D,b}^{(t), \ast})$.
            \item Compute the maximum $ m_{\max, b}^{(t),\ast} =  \max_{d = 1, \ldots, D} m_{d,b}^{(t), \ast}.$
        \end{compactenum}
        \item The regionwise $T$-year return level $\RL_{t,\mathrm{reg}}(T)$ and the return period $\mathrm{RP}_{t,\mathrm{reg}}(r)$ of an event with value $r$ can now be estimated from the empirical cumulative distribution function $\hat F_{t}^\ast$ of the sample $( m_{\max, b}^{(t),\ast})_{b=1, \dots, B}$ through
          \begin{align*}
            \widehat\RL_{t}^{\mathrm{reg}}(T) &= (\hat F_{t}^\ast)^{-1}(1-1/T),          \qquad
            \widehat{\mathrm{RP}}_{t}^{\mathrm{reg}}(r) = \frac{1}{1 - \hat F_{t}^\ast(r)}.
          \end{align*}
    \end{compactenum}
\end{algorithm}

Especially, when we have estimated the local 100-year RL, we can get an estimate of the return time this event has for the whole region. 
Likewise, when we have an estimate of the local return period of an event with value $r$, we can estimate what the event value for that return period would be for the whole region. 

We illustrate the estimators for the pooled data sets from Section \ref{sec:case}. The estimates are based on $B = 100\,000$ simulation replications and are shown in Table \ref{tab:reg_rl_loc_rl}. 
We see that the local 100-year return levels have substantially shorter region-wise return periods. In the region with 15 tiles (only cell 11 excluded), 
the estimated local 100-year RL at reference climate of 2021 can be expected to be exceeded once in approximately 19 years in at least one of the 15 tiles. We find a similar region-wise return period for the pooling region consisting of 14 tiles.  In the pooling region consisting of 11 tiles, the local 100-year return level becomes a region-wise 33-year event. This comparably larger value arises from the smaller region that is considered: the smaller the region, the less likely it is that one of the locations exceeds a high threshold. Further, as expected, we find that the region-wise 100-year return levels at reference climate of 2021 are larger than their local counterparts. For the regions consisting of 15 and 14 tiles, this increase is approximately 26\%, while it is approximately 17.3\% for the region consisting of 11 tiles.

\begin{table}[ht]
\centering
\begin{tabular}{rrrr}
 \hline
 \rule{0pt}{3ex}
 cells excluded & $ \RL_{2021}(100) $ & $ \RP_{2021}^{\mathrm{reg}}(\RL_{2021}(100))$  & $ \RL_{2021}^{\mathrm{reg}}(100)$  \\ 
  \hline
11 & 52.74 & 18.90 & 66.40 \\ 
 11, 17 & 52.82& 18.32  & 67.08  \\ 
  10, 11, 14, 15, 16 & 51.93 & 32.76 &  60.93 \\
   \hline
\end{tabular}\caption{\label{tab:reg_rl_loc_rl} Estimated local (second column) and regional (fourth column) 100-year RLs for reference climate 2021, for three possible choices of pooling regions as indicated by the first column. Column 3 shows the regional return periods of the local 100-year events. 
}
\end{table}

\subsection{A homogeneous scaling model with location-wise scaling factor} \label{subsec:if}

In this section, we maintain the temporal dynamics from the scale-GEV model from Equation \eqref{eq:ms0}. However, instead of testing for the homogeneity assumption from Equation \eqref{eq:h0}, we additionally allow for a location-wise scaling factor under the null hypothesis.
Such an approach can be useful when it is known that observations from different locations occur on different scales, but, apart from that, show a common probabilistic behaviour. In fact,
a stationary version of the following model is commonly used in hydrology, where it is known as the Index Flood approach \citep{dalrymple1960flood}.

More precisely, suppose that
\begin{align} 
\label{eq:lsm1}
M_{t,d} \sim  c_d \exp\left(\frac{\alpha \gmst(t)}{ \mu} \right) \mathrm{GEV}(\mu, \sigma, \gamma) \quad \forall t,d,
\end{align}
where $c_d>0$ is a location-specific scaling factor that we may fix to 1 at the location of primary interest (say $d=1$, i.e., $c_1=1$). 
Writing $\mu_d=c_d\mu, \sigma_d = c_d \sigma, \alpha_d = c_d \alpha$, the model in Equation \eqref{eq:lsm1} can be rewritten as
\begin{align*}
M_{t,d} \sim \mathrm{GEV}(\mu_d(t), \sigma_d(t), \gamma) \quad \forall t, d,
\end{align*}
where
\begin{align} \label{eq:ms2}
	\mu_d(t) = \mu_d \exp\left(\frac{\alpha_d \gmst(t)}{ \mu_d} \right), 
	\quad
	\sigma_d(t) = \sigma_d \exp\left(\frac{\alpha_d \gmst(t)}{ \mu_d} \right).
\end{align}
Note that the parameters $\mu_1, \dots, \mu_D, \sigma_1, \dots, \sigma_D, \alpha_1, \dots, \alpha_D$  satisfy the relationships 
\[
\frac{\mu_d}{\sigma_d} \equiv \delta, \quad \frac{\alpha_d}{\mu_d} \equiv \eta, \quad \frac{\alpha_d}{\sigma_d} \equiv \kappa
\]
for certain parameters $\delta, \eta, \kappa$; in particular, $\mu_1, \dots, \mu_D, \sigma_1, \dots, \sigma_D, \alpha_1, \dots, \alpha_D$ can be retrieved from $\mu_1, \dots, \mu_D, \delta, \eta$ (note that the constraint on $\alpha_d/\sigma_d$ is not needed, but comes as a consequence of the first two relations).
Fitting this model instead of fitting the scale-GEV distribution to each location separately has the advantage of reducing the number of parameters that need to be estimated substantially ($4+(D-1) = D +3$ instead of $4D$ parameters).
Once the local scaling factors are identified, we can bring all observations to the same scale by dividing each location by its location-specific scaling factor.

Now one can test whether such a local scaling model holds on a subset $A = \{d_1, \ldots, d_k\} \subset \{1, \ldots, D\}$ with cardinality $k = \abs{A} \geq 2$, by testing the hypothesis
\begin{align}\label{eq:h0-ls}
H_0^{\mathrm{LS}}(A): \quad \exists \,\delta_A, \eta_A,\gamma_A \,  \forall d \in A: \quad \frac{\mu_d}{\sigma_d} = \delta_A, \quad \frac{\alpha_d}{\mu_d} =\eta_A, \quad  \gamma_d = \gamma_A,
\end{align}
with a Wald-type test statistic. 
In this case, the latter is defined as 
\begin{align}\label{eq:teststat-ls}
    T_n^{\mathrm{LS}}(A) = n (g_A(\hat{\bm\theta}))^\top \left( \bm G_A(\hat{\bm\theta}) \hat{\bm\Sigma}_n \bm G_A(\hat{\bm\theta})^\top\right)^{-1}g_A(\hat{\bm\theta}),
\end{align}
where $g_A: \R^{4D} \to \R^{3(k-1)}$ is given by 
\begin{align*}
    g_A(\bm\theta) = \left( \frac{\mu_{d_1}}{\sigma_{d_1}} -  \frac{\mu_{d_2}}{\sigma_{d_2}}, \gamma_{d_1} - \gamma_{d_2},  \frac{\alpha_{d_1}}{\mu_{d_1}} -  \frac{\alpha_{d_2}}{\mu_{d_2}}, \ldots, \gamma_{d_{k-1}} - \gamma_{d_k},  \frac{\alpha_{d_{k-1}}}{\mu_{d_{k-1}}} -  \frac{\alpha_{d_k}}{\mu_{d_k}} \right)^\top, 
\end{align*}
with Jacobian matrix $G_A(\bm\theta) \in \R^{3(k-1) \times 4D}$,
since the hypothesis in Equation \eqref{eq:h0-ls} may be rewritten as $H_0^{\mathrm{LS}}(A): g_A(\bm\theta) = 0.$ 

When considering this kind of modification, the bootstrap algorithms from Section \ref{sec:boot}, steps (5)-(7), must be adapted accordingly.
In step (5), one has to estimate the parameter under the constraint of the considered null hypothesis by adapting the $\log$-likelihood accordingly. The estimated parameters are then used during the transformation step (6). Further, the test statistic in steps (6) and (7) is replaced by $T_n^{\mathrm{LS}}(A)$ from \eqref{eq:teststat-ls}.
Further details are omitted for the sake of brevity.

\subsection{General homogeneous models with smooth parametrization} \label{subsec:smooth}

In this section, we consider general GEV models in which the location, scale and shape parameters are allowed to depend in a (fixed) differentiable way on some parameter vector $\bm\vartheta \in \R^q$ and some temporal covariate $c^{(t)} \in \R^p$ with $p,q \in \mathbb N$.
More precisely, suppose that $f_\mu, f_\sigma$ and $f_\gamma$ are (known) real-valued functions of $\bm \vartheta$ and $c$ that are 
differentiable with respect to their first argument $\bm\vartheta$. We assume that, for each $d=1, \dots, d$, there exists an unknown parameter $\bm \vartheta_d$ such that $M_d^{(t)} \sim \GEV(\mu_d(t), \sigma_d(t), \gamma_d(t))$ with
\[
\mu_d(t) = f_\mu(\bm\vartheta_d; c^{(t)}), 
\quad 
\sigma_d(t) = f_\sigma(\bm\vartheta_d; c^{(t)}), 
\quad 
\gamma_d(t)= f_\gamma(\bm\vartheta_d; c^{(t)}). 
\]
The global null hypothesis of interest within this model is assumed to be expressible as 
$h(\bm\vartheta_1, \ldots, \bm\vartheta_D ) = 0$ for a differentiable function $h: \R^{qD} \to \R^s$ with $s \in \mathbb N$.

An example is given by the linear shift model that is frequently considered when modelling temperature extremes in Extreme Event Attribution studies (see \citealp{Phi20}), 
where 
\[  
\mu_d(t) = \mu_d + \alpha_d \gmst(t), 
\quad
\sigma_d(t) \equiv \sigma_d, 
\quad 
\gamma_d(t) \equiv \gamma_d.
\]
A possible global null hypothesis of interest could be
\[  
H_0: \ \exists\,  \bm\vartheta \in \R \times (0,\infty) \times \R^2 \ \forall d \in \{1, \ldots, D\}: \quad \bm\vartheta_d = \bm\vartheta,
\] 
where $\bm \vartheta=(\mu, \sigma, \gamma, \alpha)^\top$ and $\bm \vartheta_d=(\mu_d, \sigma_d, \gamma_d, \alpha_d)^\top$.

When considering this kind of extension, one has to adapt the maximum likelihood estimator as well as the estimator of its covariance matrix, hence steps (1)-(2) and (5)-(7) in the bootstrap algorithms are affected. Further details are omitted for the sake of brevity.

\section{Conclusion} 
\label{sec:conclusion}

Extreme event attribution studies can build upon a GEV scaling model. Depending on the analysed variable, it may be useful to apply spatial pooling and fit the GEV distribution to a pooled sample of observations collected at sufficiently homogeneous spatial locations as it has been done in \cite{Kre21,Tra22,Vau15}, among others.
Here, we propose several statistical methods that enable the selection of a homogeneous pooling region from a larger initial region. The BH approach was found to be quite conservative, hence some heterogeneous locations are likely to be declared homogeneous. The IM approach is a more liberal alternative with a higher proportion of rejected locations that may contain some homogeneous ones. In subsequent analyses, the selected pooling region typically results in a classical bias-variance trade-off: the larger the pooling region, the smaller the variance. At the same time, the bias may be larger, given that some heterogeneous regions may have been declared homogeneous. In practice, the tests should always be complemented by expert knowledge on the driving meteorological/climatological background processes.

To make the statistical approach to select homogeneous pooling regions for attribution studies as described here usable for the extreme event attribution community, we have developed a software package that can be freely downloaded and used by applied researchers \citep{findpoolreg}. The selection of spatial pooling regions for attribution studies may hence be based on a combination of expert knowledge and thorough statistical tests. The experts applying the methods can thereby decide between a conservative approach, which tends to reject more locations and a liberal approach which tends to accept more locations as being homogeneous. This decision depends on the a priori knowledge about the meteorology of the analysed area and the specific requirements of the study. 

If the ultimate interest is estimation of, for example, return levels, one may, as an alternative to the classical approach based on pooling selected locations, consider penalized maximum likelihood estimators with a penalty on large heterogeneities \citep{Buc21}. A detailed investigation of the resulting bias-variance trade-off would be a worthwhile topic for future research.

\section*{Declaration}

\paragraph{Ethical Approval} Not applicable.

\paragraph{Availability of supporting data} All methods are implemented in the \texttt{R} package \texttt{findpoolreg} \citep{findpoolreg} that is publicly available at \url{https://github.com/leandrazan/findpoolreg}. The data used throughout the case study is derived from the E-OBS gridded data set, publicly available at \url{https://www.ecad.eu/download/ensembles/download.php}.

\paragraph{Competing interests} The authors declare that they have no conflict of interest.

\paragraph{Funding} This work has been supported by the integrated project “Climate Change and Extreme Events - ClimXtreme Module B - Statistics (subprojects B1.2, grant number: 01LP1902B and B3.3, grant number: 01LP1902L)” funded by the German Federal Ministry of Education and Research (BMBF).

\paragraph{Authors' contributions}
LZ and AB wrote the main manuscript and worked out the mathematical details. LZ implemented all methods and carried out the Monte Carlo simulation study and the case study. FK, PL and JT contributed by extensive discussions, provided the data for the case study and improved the text.

\paragraph{Acknowledgements}
Computational infrastructure and support were provided by the Centre for Information and Media Technology at Heinrich Heine University Düsseldorf, which is gratefully acknowledged. 

\putbib
\end{bibunit}
\newpage

\begin{bibunit}
    
\numberwithin{equation}{section}
\numberwithin{figure}{section}
\numberwithin{table}{section}

\begin{center}
	
	{\Large SUPPLEMENT TO THE PAPER:  \\  ``Regional Pooling in Extreme Event Attribution Studies: an Approach Based on Multiple Statistical Testing'' }
	\vspace{.5cm}
	
	{\textsc{ Leandra Zanger, Axel B\"ucher, Frank Kreienkamp, \\ Philip Lorenz, Jordis Tradowsky}}
	
	\vspace{.28cm}

	\vspace{.28cm}

	\begin{center}
		\begin{minipage}{.6\textwidth}
			{\small \hspace{.5cm}
				This supplement contains mathematical details on the maximum likelihood estimator and the estimation of its covariance matrix, as well as additional simulation results and further details on the case study. 
References like (1) refer to equations from the main paper, while references like (A.1) or Figure B.2 refer to equations or Figures within this appendix.
				 }
		\end{minipage}
	\end{center}

\end{center}

\vspace{.5cm}




\begin{appendix}
\section{Mathematical Details} 
\label{app:mathdetails}
\subsection{Maximum Likelihood estimation}\label{app:mlest}

Throughout this section, we provide mathematical details on the coordinate-wise maximum likelihood estimator from Equation (\ref{eq:ml}). In particular, we motivate the approximate normality of $\hat {\bm \theta}=(\hat{\bm \vartheta}_1^\top, \dots, \hat{\bm \vartheta}_D^\top)^\top \in \Theta^{D}$ with mean ${\bm \theta}=({\bm \vartheta}_1^\top, \dots, {\bm \vartheta}_D^\top)^\top$ and covariance $n^{-1} \bm \Sigma_n$ with $\bm \Sigma_n = (\bm \Sigma_{n;j,k})_{j,k=1}^D \in \R^{4D \times 4D}$ as defined in Equation (\ref{eq:apprSigma}). As in the stationary GEV-model, the derivations require $\gamma>-1/2$, see \cite{BucSeg17}.

We start by some explicit formulas for the functions appearing in Equations (\ref{eq:ml}) and (\ref{eq:apprSigma}). For that purpose, let $l_{(\mu, \sigma, \gamma)}(x)$ denote the log density of the plain $\mathrm{GEV}(\mu, \sigma, \gamma)$ distribution (see Appendix B in \citealp{BucSeg17}), i.e., 
\begin{align} \label{eq:lds}
l_{(\mu, \sigma, \gamma)}(x) 
=
-\log(\sigma) - u_\gamma\Big(\frac{x-\mu}{\sigma}\Big) + (\gamma +1)\log\Big(u_\gamma \Big(\frac{x-\mu}{\sigma}\Big)\Big)
\end{align}
for $x$ such that $1+\gamma\frac{x-\mu}{\sigma}>0$; here 
\begin{align*}
u_\gamma(z) &= \begin{cases}
    (1+\gamma z)^{-\frac{1}{\gamma}},& \gamma \neq 0, \\
    \exp(-z),& \gamma = 0.
\end{cases}
\end{align*}
Then, writing $\bm \vartheta=(\mu, \sigma, \gamma, \alpha)^\top$, the $\log$-density $\ell_{\bm\vartheta}(x,c)$ from Equation~(\ref{eq:ld}) may be written as 
\begin{align} \label{eq:elll}
\ell_{\bm\vartheta}(x,c) = l_{(\mu(c), \sigma(c), \gamma)}(x), 
\end{align}
where
\begin{align*}
\mu(c) &= \mu(c\mid \mu, \alpha) = \mu\exp(\alpha c /\mu), &  
\sigma(c) &= \sigma(c \mid \mu, \sigma, \alpha) = \sigma\exp(\alpha c /\mu).
\end{align*}

We next derive formulas for the gradient and the Hessian of $\bm \vartheta\mapsto \ell_{\bm\vartheta}(x,c)$. For that purpose, let $\dot l_{(\mu,\sigma, \gamma)}(x)$ and $\ddot l_{(\mu,\sigma, \gamma)}(x)$ denote the respective gradient and Hessian of the standard GEV log density (see Appendix B in \citealp{BucSeg17}) for precise formulas). Note that, in view of the fact that the GEV distribution is a location scale family, 
\begin{align}
\dot l_{(\mu,\sigma, \gamma)}(x) 
&=  \label{eq:ls1}
T_\sigma^{-1} \dot l_{(0,1,\gamma)}\Big(\frac{x-\mu}{\sigma}\Big), 
\qquad 
T_{\sigma} = \diag( \sigma, \sigma, 1) \in \R^{3 \times 3}, \\
\ddot l_{(\mu,\sigma, \gamma)}(x) 
&=  \label{eq:ls2}
T_\sigma^{-1} \dot l_{(0,1,\gamma)}\Big(\frac{x-\mu}{\sigma}\Big) T_\sigma^{-1}. 
\end{align}
Next, consider the function $p_{c}: \Theta \to (0,\infty)^2\times \R$ defined by $p_{c}(\mu, \sigma, \gamma, \alpha) = (\mu \exp(\alpha c/\mu), \sigma\exp(\alpha c/\mu), \gamma)^\top$, whose Jacobian is given by $B_c(\mu, \sigma, \alpha)^\top$, where
\begin{align*}
    B_c(\mu, \sigma, \alpha) = \begin{pmatrix}
        \left( 1- \frac{\alpha c}{\mu} \right)\exp\left(\frac\alpha\mu c\right)&  
         -\frac{ \sigma \alpha c}{\mu^2}\exp\left(\frac\alpha\mu c\right) & 0 \\
         0 &  \exp\left(\frac\alpha\mu c\right) & 0 \\
         0 & 0 & 1 \\
         c \exp\left(\frac\alpha\mu c\right)  & 
    \frac{\sigma c}{\mu} \exp\left(\frac\alpha\mu c\right) & 0
    \end{pmatrix}.
\end{align*}
Then, in view of Equations \eqref{eq:elll} and \eqref{eq:ls1}, the multivariate chain rule yields
\begin{align}\label{eq:chainscore}
    \dot\ell_{\bm\vartheta}(x,c) 
    &=  \nonumber
    B_c(\mu, \sigma, \alpha) \cdot \dot l_{(\mu(c), \sigma(c), \gamma)}(x)  \\
    &= 
    B_c(\mu, \sigma, \alpha) \cdot T_{\sigma(c)}^{-1} \cdot \dot l_{(0,1, \gamma)}\left( \frac{x - \mu(c)}{\sigma(c)}\right).
\end{align}
In view of the multivariate product rule, this equation further implies
\begin{align}
    \ddot \ell_{\bm\vartheta}(x,c)
      &=  \nonumber
      \begin{pmatrix}
    \dot l_{(\mu(c), \sigma(c), \gamma)}(x)^\top \dot B_{c,1}(\mu, \sigma, \alpha) \\
    \dot l_{(\mu(c), \sigma(c), \gamma)}(x)^\top \dot B_{c,2}(\mu, \sigma, \alpha)  \\
   \dot l_{(\mu(c), \sigma(c), \gamma)}(x)^\top \dot B_{c,3}(\mu, \sigma, \alpha) \\
   \dot l_{(\mu(c), \sigma(c), \gamma)}(x)^\top \dot B_{c,4}(\mu, \sigma, \alpha) 
    \end{pmatrix} + 
 B_c(\mu, \sigma, \alpha) \ddot l_{(\mu(c), \sigma(c), \gamma)}(x)  B_c(\mu, \sigma, \alpha)^\top \\
    &=  \nonumber
    \begin{pmatrix} 
    \dot l_{(0,1, \gamma)}(\tfrac{ x - \mu(c)}{\sigma(c)})^\top T_{\sigma(c)}^{-1} \dot B_{c,1}(\mu, \sigma, \alpha) \\
    \dot l_{(0,1, \gamma)}(\tfrac{ x - \mu(c)}{\sigma(c)})^\top T_{\sigma(c)}^{-1} \dot B_{c,2}(\mu, \sigma, \alpha) \\
    \dot l_{(0,1, \gamma)}(\tfrac{ x - \mu(c)}{\sigma(c)})^\top T_{\sigma(c)}^{-1} \dot B_{c,3}(\mu, \sigma, \alpha) \\
    \dot l_{(0,1, \gamma)}(\tfrac{ x - \mu(c)}{\sigma(c)})^\top T_{\sigma(c)}^{-1} \dot B_{c,4}(\mu, \sigma, \alpha) \\
    \end{pmatrix}  \\
    &\hspace{1.7cm} + \label{eq:ddotl}
   B_c(\mu, \sigma, \alpha)  T_{\sigma(c)}^{-1} \ddot l_{(0,1, \gamma)}\left(\frac{ x - \mu(c)}{\sigma(c)}\right)T_{\sigma(c)}^{-1}  (B_c(\mu, \sigma, \alpha))^\top, 
\end{align}
where we used Equations \eqref{eq:ls1} and \eqref{eq:ls2} for the second equality and where $\dot B_{c,j}(\mu, \sigma, \alpha)$ denotes the Jacobian in $\R^{3 \times 4}$ of the $j$th row of $B_c(\mu, \sigma, \alpha)$ (derivative with respect to $(\mu, \sigma, \gamma, \alpha)$). The latter can be derived explicitly by a tedious but straightforward calculation; we omit precise formulas for the sake of brevity.

We next motivate the claimed normal approximation.
First of all, in view of the differentiability of $\bm \vartheta \mapsto \ell_{\bm \vartheta}$, the vector of maximum likelihood estimators is necessarily a zero of the gradient of the log-likelihood function, i.e., we have
\[
0 
= 
\frac1n \sum_{t=1}^{n} 
\begin{pmatrix}
\dot\ell_{\hat{\bm\vartheta}_1}(M_1^{(t)}, \ct) \\    
\vdots \\
\dot\ell_{\hat{\bm\vartheta}_D}(M_D^{(t)}, \ct)
\end{pmatrix}.
\]
Denoting the true parameter vector by $\bm \theta$, a Taylor expansion implies that
\begin{align*}
0 
&=
\frac1n \sum_{t=1}^{n} \begin{pmatrix}
   \dot\ell_{\bm\vartheta_1}(M_1^{(t)}, \ct) \\
    \vdots \\
 \dot\ell_{\bm\vartheta_D}(M_D^{(t)}, \ct)
\end{pmatrix}   \\
& \quad 
+ {\small \left\{ \frac1n \sum_{t=1}^{n}  \begin{pmatrix}
   \ddot\ell_{\bm\vartheta_1}(M_1^{(t)}, \ct) & 0 & \ldots & 0 \\
    0 & \ddot\ell_{\bm\vartheta_2}(M_2^{(t)}, \ct) & \ldots & 0 \\
    \vdots & \vdots & & \vdots\\
 0 & 0 & \ldots & \ddot\ell_{\bm\vartheta_D}(M_D^{(t)}, \ct)
\end{pmatrix}  \right\} \small}
(\hat {\bm \theta} - \bm \theta) + R_n \\
&\equiv L_{n, \bm \theta} + I_{n,\bm \theta}  \cdot (\hat{\bm \theta} - \bm \theta) + R_n,
\end{align*}
where $R_n$ denotes higher order terms which are negligible. 
Solving for $\sqrt n (\hat{\bm\theta} - \bm\theta) $, we obtain that
\[  
\sqrt{n} (\hat{\bm\theta} - \bm\theta) \approx 
- I_{n,\bm \theta}^{-1} \cdot \sqrt n L_{n,\bm \theta}.
\]
By Equation \eqref{eq:ddotl}, each $(4\times 4)$ block matrix $I_{n,d,\bm \vartheta_d} = \frac1n \sum_{t=1}^{n}  \ddot\ell_{\bm\vartheta_d}(M_d^{(t)}, \ct)$ on the block-diagonal of $I_{n, \bm \theta}$ is of the form
\begin{align} \label{eq:slln}
\frac1n \sum_{t=1}^n f(c^{(t)}) g(Z_d^{(t)})
\end{align}
for suitable functions $f$ and $g$,
where 
\begin{align} \label{eq:zdt}
Z_{d}^{(t)}=\{ M_d^{(t)}-\mu_d(c^{(t)})\} / \sigma_d(c^{(t)})
\end{align}
with $\mu_d(\ct) = \mu_d\exp(\alpha_d \ct /\mu_d)$ and $\sigma(\ct) =  \sigma_d\exp(\alpha_d \ct /\mu_d)$ is
$\mathrm{GEV}(0,1,\gamma_d)$-distributed and independent over $t$. Under suitable assumptions on $t \mapsto f(c^{(t)})$ (and hence on $c^{(t)}$), we obtain that the variance of $I_{n,d,\bm \vartheta_d}$ is of the order  $1/n$. As a consequence, $I_{n,d,\bm \vartheta_d}$ is close to its expectation, that is, $I_{n,d,\bm \vartheta_d} = J_{n,d,\bm \vartheta_d}+o(1)$ with $J_{n,d,\bm \vartheta_d}$ defined just after Equation (\ref{eq:apprSigma}). More precisely, in an asymptotic framework where one assumes  that $c^{(t)}=h(t/n)$ for some continuous function $h:[0,1] \to \R$, expressions as in Equation \eqref{eq:slln} converge to $\int_0^1 f(h(t)) \diff t \times \Exp[g(Z_d)]$ with $Z_d \sim \mathrm{GEV}(0,1,\gamma_d)$ (note that both the integral and the expectation exist).

Next, consider $\sqrt n L_{n,\bm \theta}$. It suffices to argue that we may apply a suitable version of the Central Limit Theorem, under suitable assumptions on $t \mapsto c^{(t)}$. Similar as for $I_{n, \bm \theta}$, by Equation \eqref{eq:chainscore}, each entry of $\sqrt n L_{n,\bm \theta}$ is of the form 
\[
\frac1{\sqrt n} \sum_{t=1}^n f(c^{(t)}) g(Z_d^{(t)})
\]
for certain functions $f$ and $g$ and for some $d\in\{1, \dots, D\}$.
In view of the independence over $t$ and the fact that $Z_{d}^{(t)} \sim \mathrm{GEV}(0,1,\gamma_d)$, we may for instance apply the Ljyapunov CLT for independent triangular arrays, see Theorem 27.3 in \cite{Bil95}. Since $\Exp[g(Z_d^{(t)})^{p}]<\infty$ for sufficiently small $p>2$ and for the functions $g$ of interest, a sufficient condition for its applicability is
\[
\lim_{n\to\infty} \frac{\sum_{t=1}^n \{ f(c^{(t)}) \}^p }{[ \sum_{t=1}^n \{f(c^{(t)}) \}^2 ]^{p/2} } = 0,
\]
which readily follows for instance if one assumes  that $c^{(t)}=h(t/n)$ for some continuous function $h:[0,1] \to \R$.

\subsection{Covariance Estimation} \label{app:covest}

Throughout this section, we provide an estimator for $\bm \Sigma_n=(\bm \Sigma_{n;j,k})_{j,k=1}^D$ defined in Equation (\ref{eq:apprSigma}). First of all, we denote by $\hat J_{n, d, \bm \vartheta_d}$ an (approximate) Hessian of the function 
\[
\bm \vartheta_d \mapsto \frac1n \sum_{t=1}^{n} 
\ell_{{\bm\vartheta}_d}(M_d^{(t)}, \ct) 
\]
evaluated at its maximizer $\hat{\bm \vartheta}_d$, possibly obtained by numerical differentiation. Note that this matrix is routinely returned by standard implementations for maximization; for instance, the \texttt{optim}-function in \texttt{R} returns an output value \texttt{hessian}.

It remains to estimate the matrix
\[
\bm C_{n,j,k} := \frac{1}{n}\sum_{t = 1}^n \Cov\big[ \dot\ell_{\bm\vartheta_j}(M_j^{(t)}, \ct),    \dot\ell_{\bm\vartheta_k}(M_k^{(t)}, \ct) \big] \in \R^{4 \times 4}
\]
for all $1\le j < k \le D$. By Equation \eqref{eq:chainscore}, we may write
\begin{multline*}
\bm C_{n,j,k}  = \frac{1}{n}\sum_{t = 1}^n B_{\ct}(\mu_j, \sigma_j, \alpha_j) T_{\sigma_j(\ct)}^{-1} \\
\times
\Cov\left(\dot l_{(0,1,\gamma_j)}(Z_j^{(t)} ),  \dot l_{(0,1,\gamma_k)}(Z_k^{(t)} )\right) T_{\sigma_k(\ct)}^{-1} ( B_{\ct}(\mu_k, \sigma_k, \alpha_k))^\top.
\end{multline*}
with $\dot l_{(0,1,\gamma)}$ the gradient of $l_{(0,1,\gamma)}$ from Equation \eqref{eq:lds} and with $Z_{j}^{(t)}$ as defined in Equation \eqref{eq:zdt}, which is $\mathrm{GEV}(0,1,\gamma_j)$-distributed for any $j=1, \dots D$.
Note that the cross  covariance $\bm \Gamma_{j,k}= \Cov(\dot l_{(0,1,\gamma_j)}(Z_j^{(t)} ),  \dot l_{(0,1,\gamma_k)}(Z_k^{(t)} ))$ does not depend on $t$, and may hence be estimated empirically after replacing the true parameters by their estimators.  More precisely, 
we obtain the estimator
\[
\hat{\bm C}_{n,j,k} 
=
\frac{1}{n}\sum_{t = 1}^n B_{\ct}(\hat \mu_j, \hat \sigma_j, \hat \alpha_j) T_{\hat \sigma_j(\ct)}^{-1} \hat {\bm \Gamma}_{n,j,k}T_{\hat \sigma_k(\ct)}^{-1}  B_{\ct}(\hat \mu_k, \hat \sigma_k, \hat \alpha_k)
\]
where $\hat\sigma_j(c) = \hat\sigma_j\exp(\hat\alpha_j c /\hat\mu_j)$ and where $\hat {\bm \Gamma}_{n,j,k}$ denotes the empirical cross  covariance matrix of the two samples
$( \dot l_{(0, 1, \hat\gamma_j)}(\hat Z_j^{(t)}))_{t=1}^n$  and 
$( \dot l_{(0, 1, \hat\gamma_k)}(\hat Z_k^{(t)}))_{t=1}^n $ with
\[
\hat Z_j^{(t)} = \frac{ M_j^{(t)} - \hat\mu_j(\ct)}{\hat\sigma_j(\ct)}
\]
and $\hat\mu_j(c) = \hat\mu_j\exp(\hat\alpha_j c /\hat\mu_j)$. The final estimator for $\bm \Sigma_n$ is
$
\hat {\bm \Sigma}_n = (\hat {\bm \Sigma}_{n;j,k})_{j,k=1}^D
$
with
\[
\hat {\bm \Sigma}_{n;j,k} = \hat J_{n, j, \bm \vartheta_d}^{-1} \hat{\bm C}_{n,j,k}  \hat J_{n, k, \bm \vartheta_k}^{-1}.
\]

\section{Additional results of the simulation study}\label{app:sim}

\subsection{Additional results for record length $n = 75$}

\begin{figure}[tbh]
      \makebox[\textwidth][c]{
        \includegraphics[width = 1.1\textwidth]{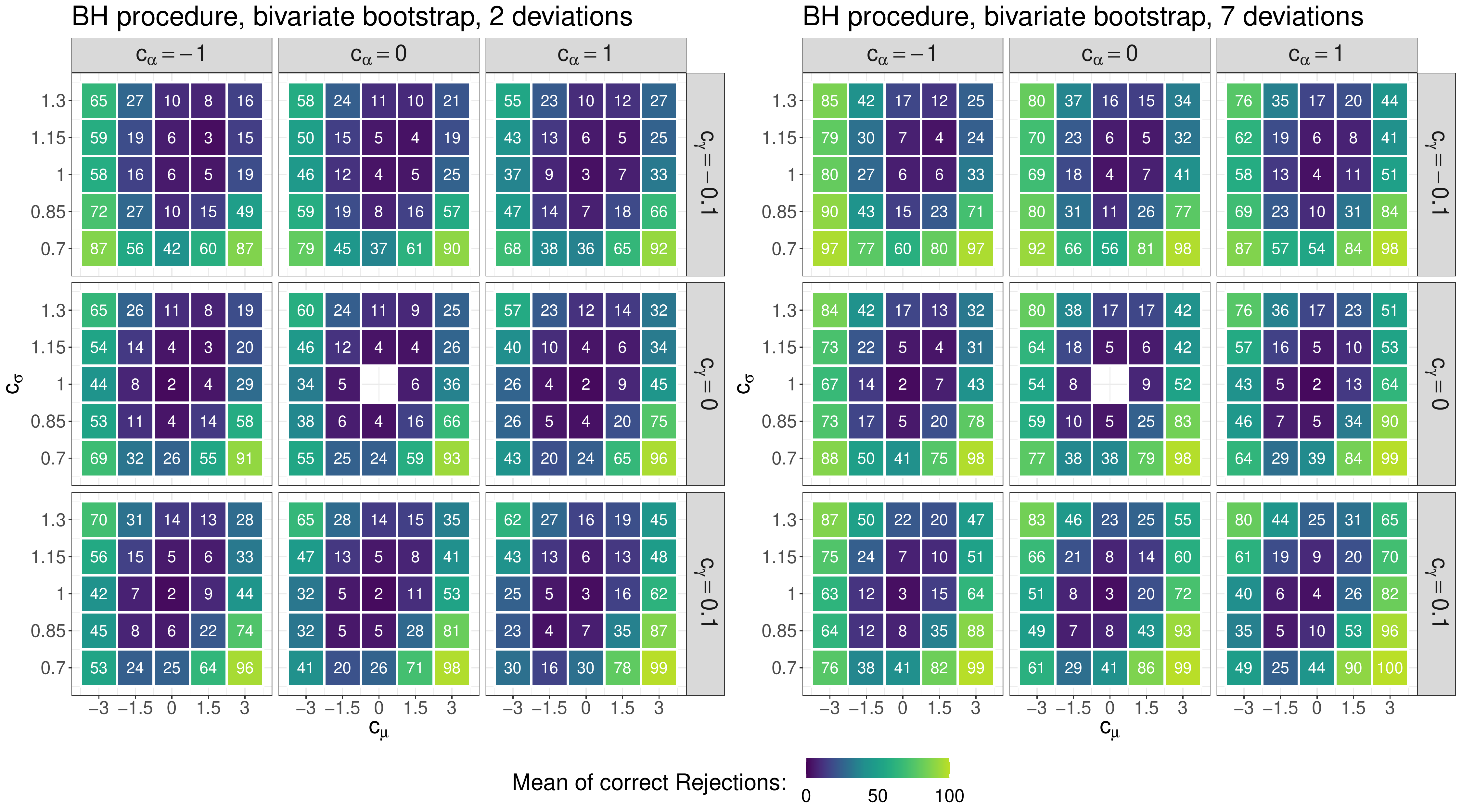}
        }
        \caption{Proportion of correct rejections in $\%$ obtained with the Benjamini Hochberg procedure at a level of 0.1, in the setting where two stations deviate from the rest (left column) or 7 stations deviate from the rest (right column), with the bootstrap procedure based on bivariate extreme value distributions.
        The axis and facets are as described in Figure \ref{fig:power_ms_on16}.
}
        \label{fig:power-biv-bh}
    \end{figure}

\begin{figure}[tbh]
      \makebox[\textwidth][c]{
        \includegraphics[width = 1.1\textwidth]{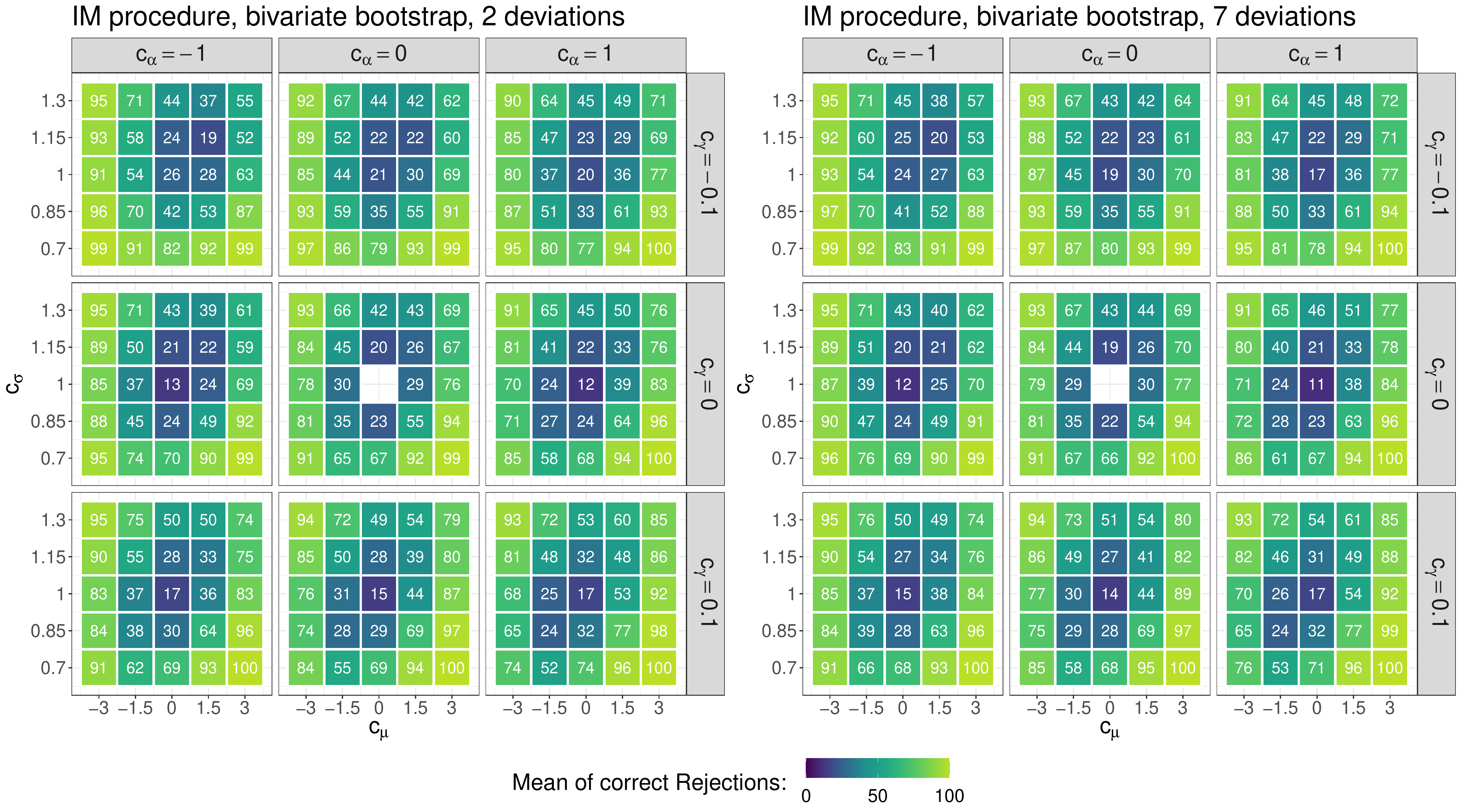}
        }
        \caption{Proportion of correct rejections in $\%$ obtained with the Benjamini Hochberg procedure at a level of 0.1, in the setting where two stations deviate from the rest (left column) or 7 stations deviate from the rest (right column), with the bootstrap procedure based on bivariate extreme value distributions. The axis and facets are as described in Figure \ref{fig:power_ms_on16}.}
        \label{fig:power-biv-im}
    \end{figure}

\begin{figure}[tbh]
      \makebox[\textwidth][c]{
        \includegraphics[width = 1.1\textwidth]{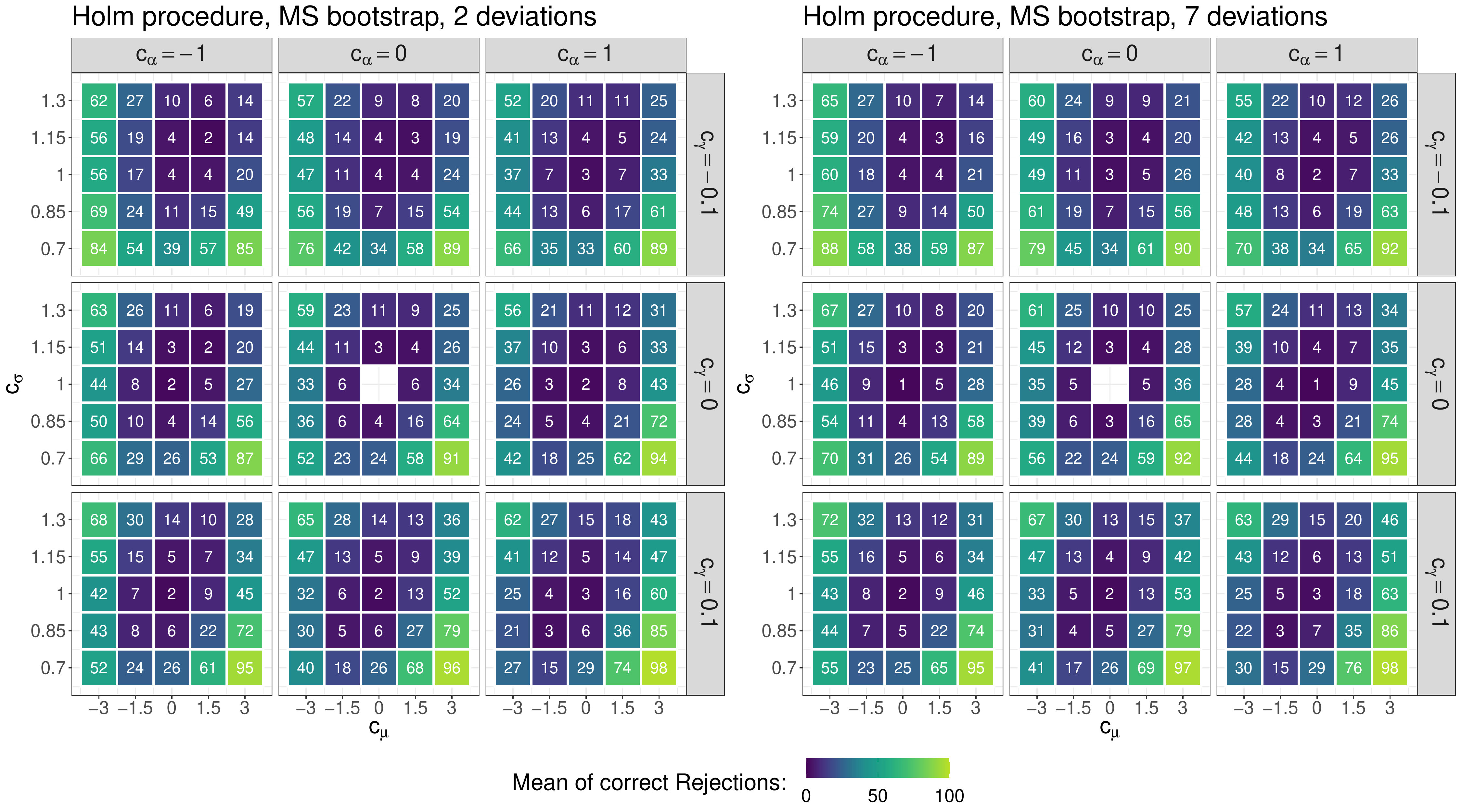}
        }
        \caption{Proportion of correct rejections in $\%$ obtained with the Holm procedure at a level of 0.1, in the setting where two stations deviate from the rest (left column) or 7 stations deviate from the rest (right column), with the bootstrap procedure based on max-stable processes. The axis and facets are as described in Figure \ref{fig:power_ms_on16}.}
        \label{fig:power-ms-holm}
    \end{figure}

\begin{figure}[tbh]
      \makebox[\textwidth][c]{
        \includegraphics[width = 1.1\textwidth]{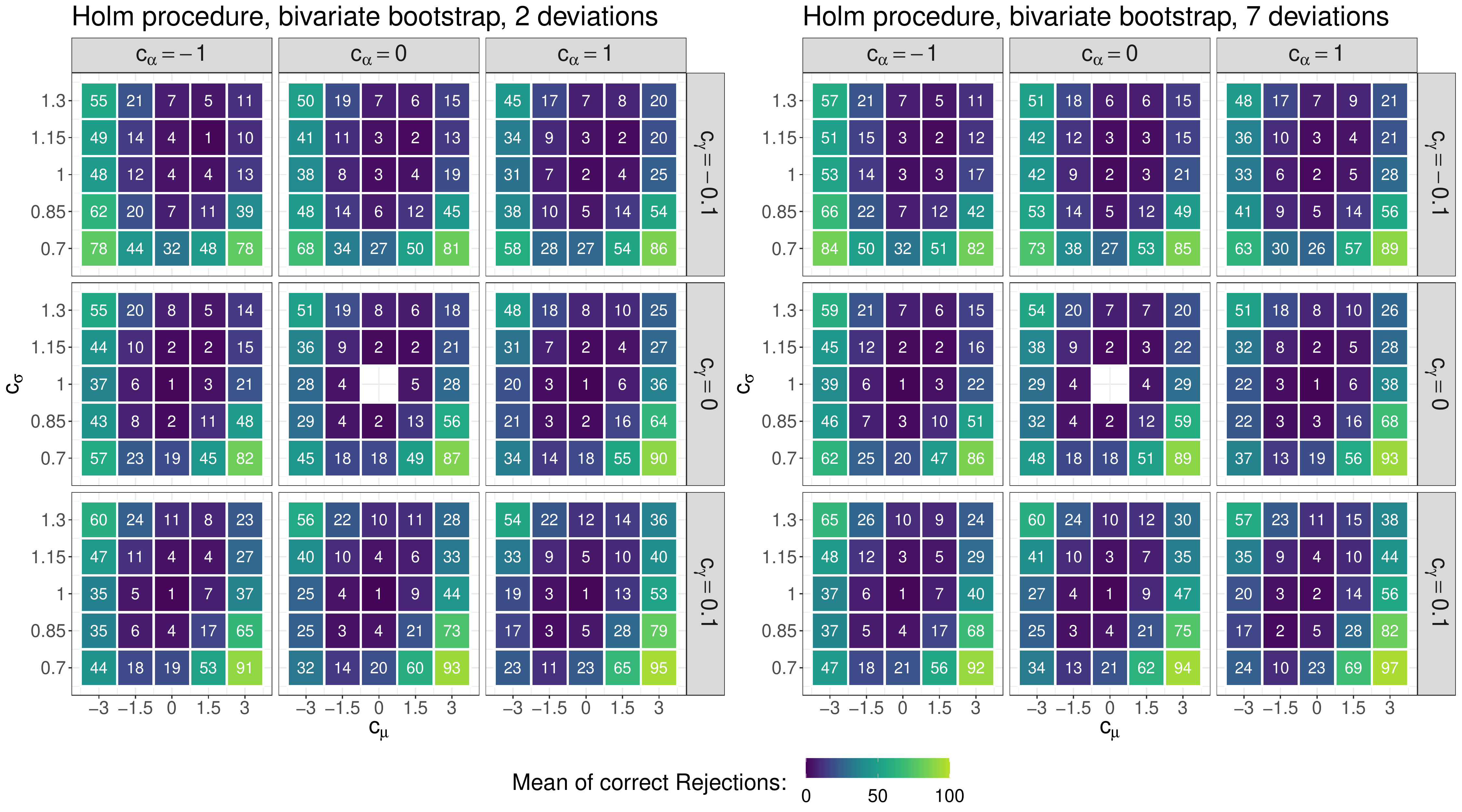}
        }
        \caption{Proportion of correct rejections in $\%$ obtained with the Holm procedure at a level of 0.1, in the setting where two stations deviate from the rest (left column) or 7 stations deviate from the rest (right column), with the bootstrap procedure based on bivariate extreme value distributions. The axis and facets are as described in Figure \ref{fig:power_ms_on16}.}
        \label{fig:power-biv-holm}
    \end{figure}

    We report the power properties obtained with the BH and IM method for procedure (B3) in Figures \ref{fig:power-biv-bh} and \ref{fig:power-biv-im}, respectively. 
   Power properties obtained with the Holm method are shown in Figures \ref{fig:power-ms-holm} (for (B2)) and \ref{fig:power-biv-holm} (for (B3)).


\subsection{Additional results for record length $n = 100$}

\begin{figure}[tbh]
\makebox[\textwidth][c]{
    \includegraphics[width=1.1\textwidth]{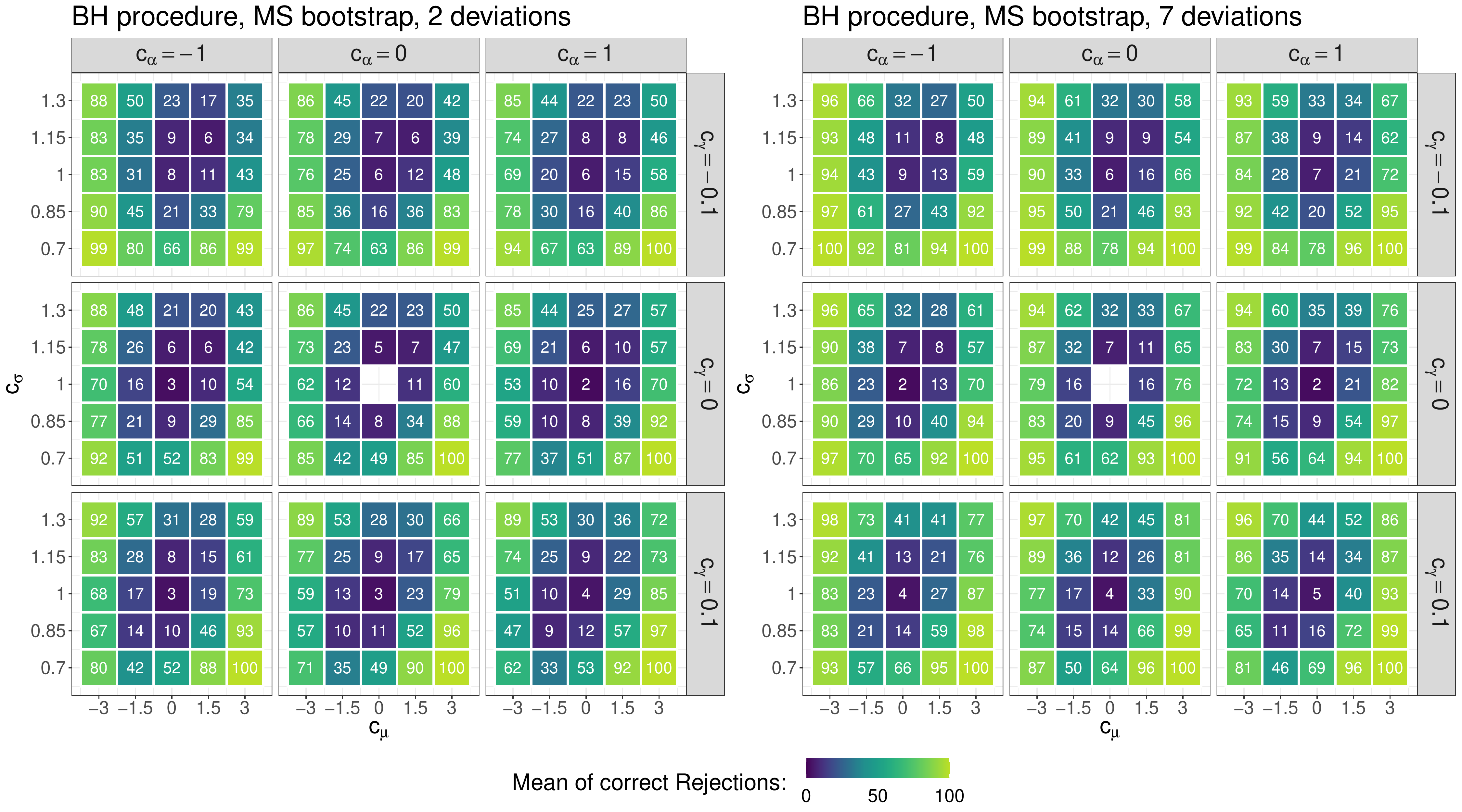}
}	
\caption{Proportion of correct rejections in $\%$ obtained with the Benjamini Hochberg procedure at a level of 0.1, in the setting where two stations deviate from the rest (left column) or 7 stations deviate from the rest (right column), with the bootstrap procedure based on max-stable processes and record length $n=100$. The axis and facets are as described in Figure \ref{fig:power_ms_on16}.} 
\label{fig:bh-power-n100}
\end{figure}

\begin{figure}[tbh]
\makebox[\textwidth][c]{
    \includegraphics[width=1.1\textwidth]{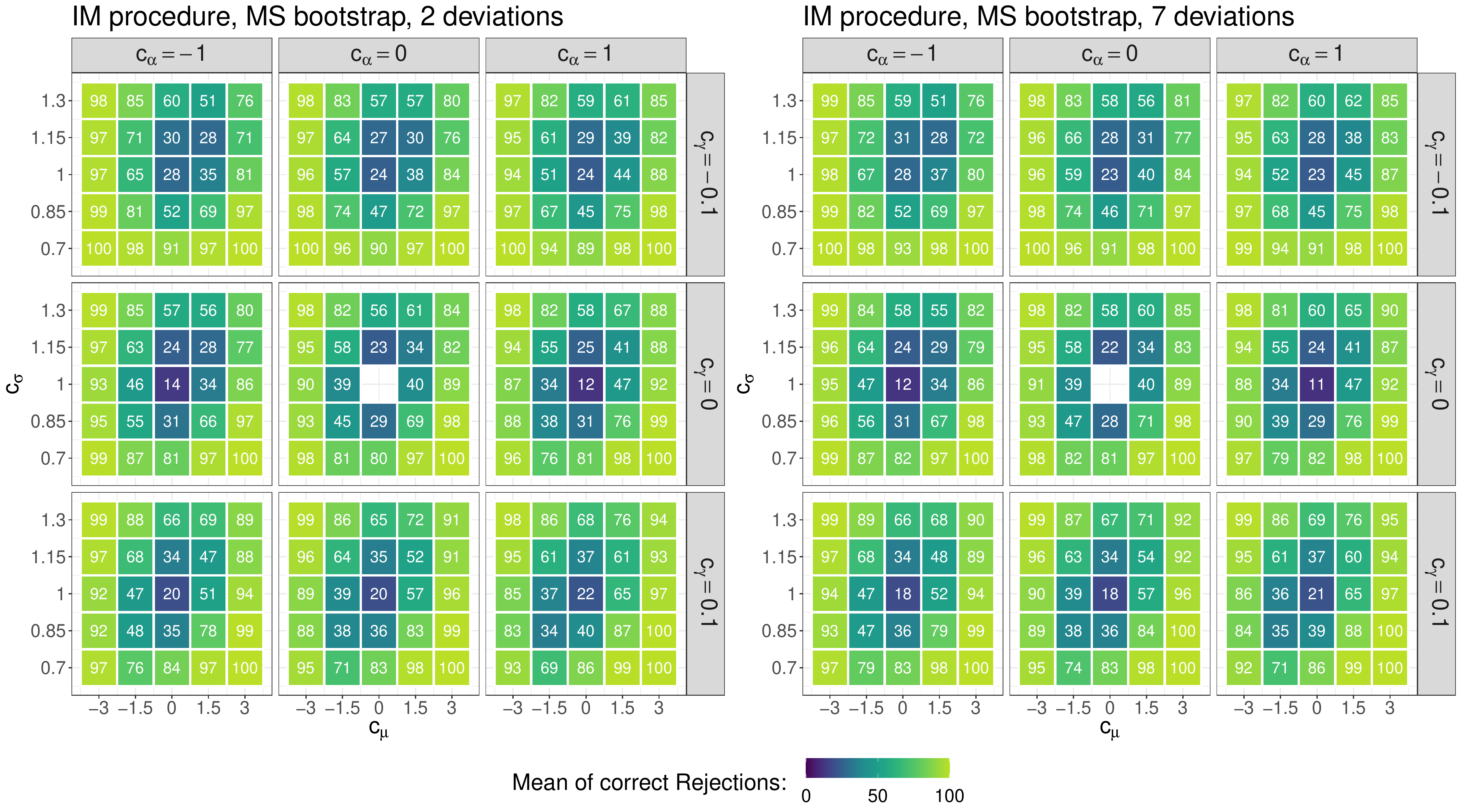}
}	
\caption{Proportion of correct rejections in $\%$ obtained when ignoring the multiple testing problem,at a level of 0.1, in the setting where two stations deviate from the rest (left column) or 7 stations deviate from the rest (right column), with the bootstrap procedure based on max-stable processes and record length $n=100$. The axis and facets are as described in Figure \ref{fig:power_ms_on16}.} 
\label{fig:im-power-n100}
\end{figure}

Since the bootstrap procedures implicitly depend on the asymptotic distribution of the test statistic, we repeated the simulation study with a larger sample size, in order to investigate the sample size's impact on the performance of the bootstrap procedure. The location-wise sample size is increased to $n =100$. Again, the FDR and FWER are reported (Table \ref{tab:fdr-n100}), as well as the power plots for BH and (B2) in Figure~\ref{fig:bh-power-n100}, and for IM and (B2) in Figure~\ref{fig:im-power-n100}. As expected, the error rates are again sufficiently controlled by those methods that claim to do so, while the power has substantially increased (on average by 50\% for the BH method and by 18\% for the IM method).
The results for the other methods and (B3) were again similar.

\begin{table}[t]
\centering
\makebox[\textwidth][c]{
\footnotesize
\begin{tabular}{l|| rr|rr|rr || rr|rr|rr}
\hline 
Method & \multicolumn{2}{c|}{min FDR} & \multicolumn{2}{c|}{max FDR} & \multicolumn{2}{c||}{mean FDR} & \multicolumn{2}{c|}{min FWER} & \multicolumn{2}{c|}{max FWER} & \multicolumn{2}{c}{mean FWER}  \\ 
& (B2) & (B3) & (B2) & (B3) & (B2) & (B3) & (B2) & (B3) & (B2) & (B3) & (B2) & (B3)  \\ 
  \hline \hline \addlinespace[.2cm]
  \multicolumn{13}{l}{\quad\textit{Scenario 1: $|A_\dev|=2$}} \\ \hline
  \hline
BH & 6.9 & 5.2 & 12.4 & 11.4 & 9.4 & 7.8 & 8.7 & 6.8 & 23.9 & 20.4 & 15.8 & 13.4 \\ 
  Holm & 2.7 & 1.4 & 10.3 & 8.0 & 6.4 & 4.6 & 6.8 & 3.9 & 12.9 & 9.4 & 9.5 & 6.7 \\ 
 IM & 26.4 & 25.7 & 60.3 & 59.9 & 35.2 & 34.6 & 53.4 & 52.7 & 64.4 & 63.8 & 59.2 & 58.3 \\ 

       \hline  \addlinespace[.2cm]
  \multicolumn{13}{l}{\quad\textit{Scenario 2: $|A_\dev|=7$}} \\  \hline
  
 BH & 4.0 & 2.7 & 10.7 & 8.3 & 5.6 & 5.0 & 6.3 & 5.1 & 32.4 & 32.6 & 21.6 & 19.9 \\ 
  Holm & 0.8 & 0.5 & 10.3 & 6.5 & 2.9 & 2.1 & 4.2 & 3.0 & 11.2 & 8.0 & 7.1 & 5.2 \\ 
  IM & 8.2 & 7.8 & 60.1 & 59.7 & 14.2 & 13.8 & 41.8 & 40.9 & 60.1 & 59.7 & 47.9 & 46.6 \\ 
   \hline 
\end{tabular}
}
\caption{False Discovery Rate (FDR) and Familiywise Error Rate (FWER) for the three p-value combination methods from Section~\ref{subsec:combineTeststats} and the two bootstrap methods (B2) and (B3), obtained in the simulations with record length $n =100$. The stated values are the minimum, maximum and mean across the 224 models from each scenario. 
} \label{tab:fdr-n100}
\end{table}

\section{Additional results for the case study}\label{app:case}

The complete results of the bootstrap procedures applied to the $4\times 4$ can be found in Table~\ref{tab:p-case-16}. For the $6\times 6$ grid, 
the complete results of the bootstrap based on bivariate extreme value distributions can be found in Table~\ref{tab:p-case-biv}, and the results for the bootstrap procedure based on max-stable processes in Table~\ref{tab:p-case-ms}.

\begin{figure}[tbh]
  \makebox[\textwidth][c]{
  	\includegraphics[width=1.2\textwidth]{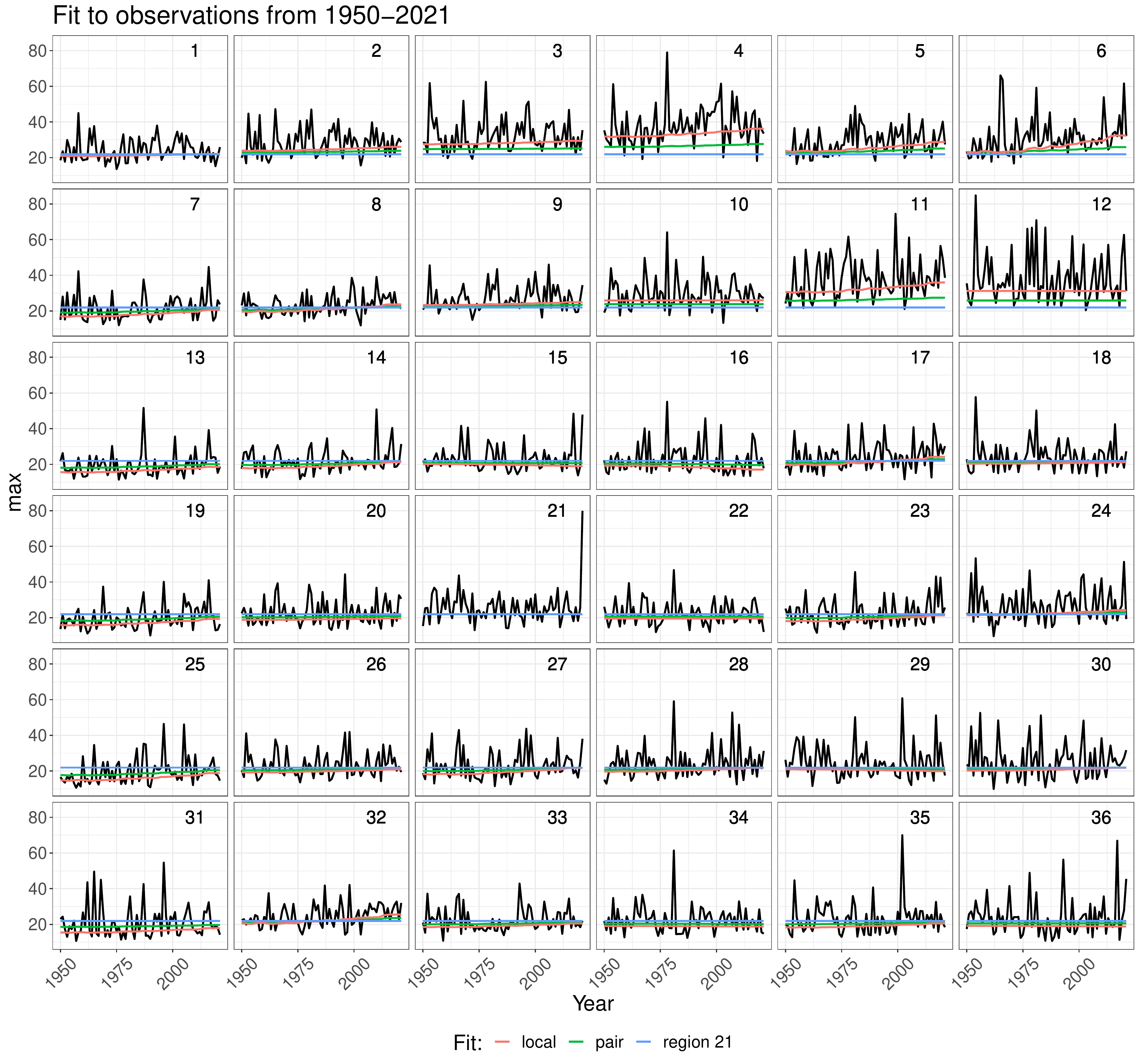}	}\caption{
   \label{fig:obs_with_trend} Observations and fitted trend as estimated for each location $d, \, d = 1, \ldots, 36,$ (red line) as well as the trend obtained from the GEV fit for location 21 (blue line)
   and the GEV-fit obtained from
   the pooled sample consisting of the pair $(21, d), \, d = 1, \ldots, 36$ (green line). Location labels are given in the top right corner. }
  \end{figure}

The time series used throughout the case study are shown in Figure~\ref{fig:obs_with_trend}. Along with the Apr-Sep maxima of 1950-2021, we plot values of 
$$\hat\mu(t) = \hat\mu\exp\left( \frac{ \hat\alpha\gmst(t)}{\hat\mu}  \right),$$
where $\hat \mu $ and $\hat\alpha$ are estimated on the data of
location 21 only (blue line), the respective location $d$ (red line) or the pooled data of the pair $(21, d)$ (green line), for $ d \in \{ 1, \ldots, 36\} \setminus\{21\}$.
Note that these three lines should not differ much when the homogeneity assumption holds. On the other hand, perfectly matching lines do not imply that the homogeneity assumption is true, since they do not give any information about the scale and shape parameter of the distributions.

\begin{table}[ht]
\centering
\begin{tabular}{l r | rrr || rrr}
  \hline
   &  &\multicolumn{3}{c||}{MS bootstrap (B2)} &  \multicolumn{3}{c }{bivariate bootstrap (B3)} \\
   \hline
Pair &  $t_n$ & $p_{\text{raw}}$ & $p_{\text{BH}}$  & $p_{\text{Holm}}$  & $p_{\text{raw}}$ & $p_{\text{BH}}$  & $p_{\text{Holm}}$ \\ 
  \hline
(21, 11) & 78.8 & \bf 0.00 & \bf 0.00 &\bf  0.00 & \bf 0.05 & \bf 0.75 & \bf 0.75 \\ 
  (21, 16) & 15.6 & \bf 1.60 & 10.64 & 22.39 & \bf 1.45 & \bf 9.90 & 20.29 \\ 
  (21, 15) & 15.2 & \bf 2.50 & 10.64 & 32.48 & \bf 2.50 & \bf 9.90 & 32.48 \\ 
  (21, 10) & 13.6 & \bf 3.40 & 10.64 & 40.78 & \bf 3.30 & \bf 9.90 & 36.58 \\ 
  (21, 14) & 13.7 & \bf 3.55 & 10.64 & 40.78 & \bf 3.05 & \bf 9.90 & 36.58 \\ 
  (21, 23) & 12.1 & \bf 5.30 & 13.24 & 52.97 & \bf 6.30 & 15.74 & 62.97 \\ 
  (21, 20) & 10.4 & \bf 7.15 & 15.09 & 64.32 & \bf 9.85 & 16.36 & 78.76 \\ 
  (21, 27) & 10.6 & \bf 8.05 & 15.09 & 64.37 & \bf 8.10 & 16.36 & 72.86 \\ 
  (21, 9) & 9.8 & \bf 10.00 & 15.59 & 69.97 & 10.44 & 16.36 & 78.76 \\ 
  (21, 22) & 9.4 & 10.39 & 15.59 & 69.97 & 11.69 & 16.36 & 78.76 \\ 
  (21, 8) & 9.0 & 13.04 & 17.79 & 69.97 & 11.99 & 16.36 & 78.76 \\ 
  (21, 26) & 7.2 & 20.34 & 25.42 & 81.36 & 22.04 & 27.55 & 88.16 \\ 
  (21, 17) & 4.4 & 46.08 & 53.17 & 100.00 & 45.73 & 52.76 & 100.00 \\ 
  (21, 28) & 3.9 & 52.17 & 55.90 & 100.00 & 50.92 & 54.56 & 100.00 \\ 
  (21, 29) & 2.8 & 66.92 & 66.92 & 100.00 & 65.77 & 65.77 & 100.00 \\ 
   \hline
\end{tabular}\caption{(Adjusted) p-values for all three methods from Section~\ref{subsec:combineTeststats} obtained with both bootstrap methods applied to the $4\times 4$ grid. Values that are significant at the 10\%-level are in boldface. Results are based on $B = 2000$ bootstrap replications. }
\label{tab:p-case-16}
\end{table}

\begin{table}[ht]
\centering
\begin{tabular}{lrrrr}
  \hline
Pair &  $t_n$ & $p_{\text{raw}}$ & $p_{\text{BH}}$   & $p_{\text{Holm}}$  \\ 
  \hline
(21, 3) & 41.5 & \bf 0.10 & \bf0.50  & \bf3.50 \\ 
  (21, 4) & 87.3 &\bf 0.10 & \bf0.50  & \bf3.50 \\ 
  (21, 6) & 46.0 &\bf 0.10 &\bf 0.50  & \bf3.50 \\ 
  (21, 11) & 78.8 & \bf0.10 & \bf0.50  & \bf3.50 \\ 
  (21, 12) & 75.2 & \bf0.10 & \bf0.50  & \bf3.50 \\ 
  (21, 13) & 33.7 & \bf0.10 & \bf0.50  & \bf3.50 \\ 
  (21, 25) & 38.5 &\bf 0.10 &\bf 0.50 & \bf3.50 \\ 
  (21, 19) & 28.6 & \bf 0.20 &\bf 0.87  & \bf5.59 \\ 
  (21, 31) & 33.3 & \bf0.30 & \bf1.17 & \bf8.09 \\ 
  (21, 7) & 16.1 & \bf1.50 &\bf 5.24  & 38.96 \\ 
  (21, 5) & 15.7 & \bf1.70 &\bf 5.40 & 42.46 \\ 
  (21, 16) & 15.6 & \bf 2.00 & \bf5.83  & 47.95 \\ 
  (21, 10) & 13.6 & \bf2.70 &\bf 6.99 & 62.04 \\ 
  (21, 15) & 15.2 &\bf 2.80 & \bf6.99  & 62.04 \\ 
  (21, 14) & 13.8 &\bf 4.10 & \bf9.56  & 86.01 \\ 
  (21, 33) & 12.5 &\bf 4.70 & 10.27  & 93.91 \\ 
  (21, 23) & 12.1 &\bf 6.89 & 14.19  & 100.00 \\ 
  (21, 20) & 10.4 &\bf 8.39 & 16.19  & 100.00 \\ 
  (21, 27) & 10.6 & \bf8.79 & 16.19  & 100.00 \\ 
  (21, 36) & 10.5 &\bf 9.89 & 16.98  & 100.00 \\ 
  (21, 34) & 9.9 & 10.19 & 16.98  & 100.00 \\ 
  (21, 2) & 9.6 & 11.19 & 17.03 & 100.00 \\ 
  (21, 9) & 9.8 & 11.19 & 17.03 & 100.00 \\ 
  (21, 22) & 9.4 & 13.69 & 19.44 & 100.00 \\ 
  (21, 8) & 9.0 & 13.89 & 19.44  & 100.00 \\ 
  (21, 35) & 8.2 & 15.88 & 21.38 & 100.00 \\ 
  (21, 26) & 7.2 & 19.78 & 25.64 & 100.00 \\ 
  (21, 30) & 6.1 & 29.17 & 36.46  & 100.00 \\ 
  (21, 32) & 5.8 & 33.07 & 39.91  & 100.00 \\ 
  (21, 17) & 4.4 & 46.75 & 54.55 & 100.00 \\ 
  (21, 28) & 3.9 & 52.75 & 59.55  & 100.00 \\ 
  (21, 29) & 2.8 & 66.13 & 72.33  & 100.00 \\ 
  (21, 24) & 2.7 & 70.23 & 74.49  & 100.00 \\ 
  (21, 1) & 2.4 & 73.13 & 75.28  & 100.00 \\ 
  (21, 18) & 1.8 & 83.82 & 83.82 & 100.00 \\ 
   \hline
\end{tabular} \caption{(Adjusted) p-values for all three methods from Section~\ref{subsec:combineTeststats} obtained with the bootstrap based on bivariate extreme value distributions. Values that are significant at the 10\%-level are in boldface. Results are based on $B = 2000$ bootstrap replications. }
\label{tab:p-case-biv}
\end{table}

\begin{table}[ht]
\centering
\begin{tabular}{lrrrr}
  \hline
 Pair &  $t_n$ & $p_{\text{raw}}$ & $p_{\text{BH}}$    & $p_{\text{Holm}}$ \\ 
  \hline
(21, 3) & 41.5 & \bf 0.00 & \bf 0.00 &  \bf 0.00 \\ 
  (21, 4) & 87.2 &\bf 0.00 & \bf0.00 & \bf 0.00 \\ 
  (21, 11) & 78.8 & \bf0.00 &\bf 0.00 &\bf0.00 \\ 
  (21, 12) & 75.2 & \bf0.00 & \bf0.00 & \bf 0.00 \\ 
  (21, 25) & 38.5 &\bf 0.00 & \bf0.00  &\bf 0.00 \\ 
  (21, 6) & 46.0 &\bf 0.10 & \bf0.50 & \bf3.00 \\ 
  (21, 13) & 33.4 &\bf 0.10 & \bf0.50 & \bf3.00 \\ 
  (21, 31) & 33.3 & \bf0.20 & \bf0.87  &\bf 5.59 \\ 
  (21, 19) & 28.7 & \bf0.30 &\bf 1.17  & \bf8.09 \\ 
  (21, 7) & 16.0 & \bf1.80 & \bf6.29  & 46.75 \\ 
  (21, 5) & 15.7 & \bf2.60 &\bf 7.53 & 64.94 \\ 
  (21, 15) & 15.2 & \bf2.70 & \bf7.53 & 64.94 \\ 
  (21, 16) & 15.6 & \bf2.80 &\bf 7.53  & 64.94 \\ 
  (21, 14) & 13.7 & \bf3.70 & \bf9.24  & 81.32 \\ 
  (21, 10) & 13.6 & \bf4.40 & 10.26  & 92.31 \\ 
  (21, 23) & 12.1 & \bf5.09 & 11.15  & 100.00 \\ 
  (21, 33) & 12.5 & \bf5.89 & 12.13  & 100.00 \\ 
  (21, 27) & 10.6 & \bf8.59 & 15.96 & 100.00 \\ 
  (21, 20) & 10.4 & \bf8.89 & 15.96 & 100.00 \\ 
  (21, 36) & 10.5 & \bf9.19 & 15.96 & 100.00 \\ 
  (21, 9) & 9.8 & \bf9.59 & 15.96  & 100.00 \\ 
  (21, 2) & 9.6 & 10.49 & 15.96  & 100.00 \\ 
  (21, 34) & 9.9 & 10.49 & 15.96  & 100.00 \\ 
  (21, 22) & 9.4 & 10.99 & 16.03  & 100.00 \\ 
  (21, 8) & 9.0 & 12.39 & 17.34  & 100.00 \\ 
  (21, 35) & 8.2 & 16.38 & 22.05 & 100.00 \\ 
  (21, 26) & 7.2 & 21.88 & 28.36 & 100.00 \\ 
  (21, 30) & 6.1 & 28.77 & 35.96 & 100.00 \\ 
  (21, 32) & 5.8 & 31.87 & 38.46  & 100.00 \\ 
  (21, 17) & 4.4 & 46.25 & 53.96 & 100.00 \\ 
  (21, 28) & 3.9 & 50.95 & 57.52 & 100.00 \\ 
  (21, 29) & 2.8 & 66.33 & 70.99  & 100.00 \\ 
  (21, 24) & 2.7 & 66.93 & 70.99 & 100.00 \\ 
  (21, 1) & 2.4 & 71.03 & 73.12  & 100.00 \\ 
  (21, 18) & 1.8 & 81.92 & 81.92  & 100.00 \\ 
   \hline
\end{tabular} \caption{(Adjusted) p-values for all three methods from Section~\ref{subsec:combineTeststats} obtained with the bootstrap based on max-stable processes. Values that are significant at the 10\%-level are in boldface. Results are based on $B = 2000$ bootstrap replications.}
 \label{tab:p-case-ms}
\end{table}

\end{appendix}

\clearpage
\putbib
\end{bibunit}


\end{document}